\documentclass[useAMS,usenatbib]{mnras}
\usepackage{natbib}
\usepackage[english]{babel}
\usepackage{graphicx,amsmath,amsfonts,amssymb}
\usepackage{colortbl}
\usepackage{tabu,booktabs}
\usepackage{xcolor}
\usepackage{xspace}
\usepackage{txfonts}
\usepackage{mathrsfs,bm}

\def \etal {et~al.~}

\newcommand{\hMpc}{{\ifmmode{h^{-1}{\rm Mpc}}\else{$h^{-1}$Mpc}\fi}}
\newcommand{\Mpc}{{\ifmmode{{\rm Mpc}}\else{Mpc}\fi}}
\newcommand{\hkpc}{{\ifmmode{h^{-1}{\rm kpc}}\else{$h^{-1}$kpc}\fi}}
\newcommand{\kpc}{{\ifmmode{ {\rm kpc} }\else{{\rm kpc}}\fi}}
\newcommand{\kms}{{\ifmmode{ {\rm km\,s^{-1}} }\else{ ${\rm km\,s^{-1}}$ }\fi}}
\newcommand{\hMsun}{{\ifmmode{h^{-1}{\rm {M_{\odot}}}}\else{$h^{-1}{\rm{M_{\odot}}}$}\fi}}
\newcommand{\Msun}{{\ifmmode{{\rm M}_{\odot}}\else{${\rm M}_{\odot}$}\fi}}
\newcommand{\Mhalo}{{\ifmmode{M_{\rm halo}}\else{$M_{\rm halo}$}\fi}}
\newcommand{\Rvir}{{\ifmmode{R_{\rm vir}}\else{$R_{\rm vir}$}\fi}}
\newcommand{\Mstar}{{\ifmmode{M_{\rm star}}\else{$M_{\rm star}$}\fi}}
\newcommand{\Vrot}{{\ifmmode{V_{\rm rot}}\else{$V_{\rm rot}$}\fi}}
\newcommand{\ltsima}{$\; \buildrel < \over \sim \;$}
\newcommand{\gtsima}{$\; \buildrel > \over \sim \;$}
\newcommand{\lsim}{\lower.5ex\hbox{\ltsima}}
\newcommand{\gsim}{\lower.5ex\hbox{\gtsima}}

\def\lesssim{\mathrel{\hbox{\rlap{\hbox{\lower4pt\hbox{$\sim$}}}\hbox{$<$}}}}
\def\gtrsim{\mathrel{\hbox{\rlap{\hbox{\lower4pt\hbox{$\sim$}}}\hbox{$>$}}}}

\newcommand{\Sec}[1]{Section~\ref{#1}}

\newcommand{\beq}{\begin{equation}}
\newcommand{\eeq}{\end{equation}}
\def\beqa{\begin{eqnarray}}
\def\eeqa{\end{eqnarray}}
\def\LCDM{\ensuremath{\Lambda}CDM}

\def\head{ \vbox to 0pt{\vss \hbox to 0pt{\hskip 440pt\rm
      LA-UR-10-07069\hss} \vskip 25pt}}

\newcommand{\mat}{\ensuremath{\mathbfss}}
\newcommand{\bnabla}{\bm{\nabla}}

\newcommand{\eps}{\varepsilon}
\newcommand{\CR}{\mathrm{cr}}
\newcommand{\bvel}{\ensuremath{\bm{\varv}}}
\newcommand{\bB}{\ensuremath{\mathbfit{B}}}
\newcommand{\bb}{\ensuremath{\mathbfit{b}}}
\newcommand{\bcdot}{\ensuremath{%
  \mathchoice%
   {\mskip\thinmuskip\lower0.2ex\hbox{\scalebox{1.5}{$\cdot$}}\mskip\thinmuskip}}%
   {\mskip\thinmuskip\lower0.2ex\hbox{\scalebox{1.5}{$\cdot$}}\mskip\thinmuskip}%
   {\lower0.3ex\hbox{\scalebox{1.2}{$\cdot$}}}%
   {\lower0.3ex\hbox{\scalebox{1.2}{$\cdot$}}}%
}

\def \kms {\ifmmode  \,\rm km\,s^{-1} \else $\,\rm km\,s^{-1}  $ \fi }
\def \kpc {\ifmmode  {\rm kpc}  \else ${\rm  kpc}$ \fi  }  
\def \hkpc {\ifmmode  {h^{-1}\rm kpc}  \else ${h^{-1}\rm kpc}$ \fi  }  
\def \hMpc {\ifmmode  {h^{-1}\rm Mpc}  \else ${h^{-1}\rm Mpc}$ \fi  }  
\def \Mpch {\ifmmode  {h^{-1}\rm Mpc}  \else ${h^{-1}\rm Mpc}$ \fi  }  
\def \Msun {\ifmmode {\rm M}_{\odot} \else ${\rm M}_{\odot}$ \fi} 
\def \hMsun {\ifmmode h^{-1}\,\rm M_{\odot} \else $h^{-1}\,\rm M_{\odot}$ \fi}

\def \LCDM {\ifmmode \Lambda{\rm CDM} \else $\Lambda{\rm CDM}$ \fi}
\def \sig8 {\ifmmode \sigma_8 \else $\sigma_8$ \fi} 
\def \OmegaM {\ifmmode \Omega_{\rm m} \else $\Omega_{\rm m}$ \fi} 
\def \Omegab {\ifmmode \Omega_{\rm b} \else $\Omega_{\rm b}$ \fi} 
\def \OmegaL {\ifmmode \Omega_{\rm \Lambda} \else $\Omega_{\rm \Lambda}$\fi} 
\def \Deltavir {\ifmmode \Delta_{\rm vir} \else $\Delta_{\rm vir}$ \fi}
\def \rhocrit {\ifmmode \rho_{\rm crit} \else $\rho_{\rm crit}$ \fi}
\def \rhou {\ifmmode \rho_{\rm u} \else $\rho_{\rm u}$ \fi}
\def \zc {\ifmmode z_{\rm c} \else $z_{\rm c}$ \fi}

\def\aj{AJ}%
\def\araa{ARA\&A}%
\def\apj{ApJ}%
\def\apjl{ApJ}%
\def\aap{A\&A}%
\def\jcap{J. Cosmology Astropart. Phys.}%
\def\mnras{MNRAS}%
\def\prd{Phys.~Rev.~D}%
\def\pasp{PASP}%
\def\ssr{Space~Sci.~Rev.}%
\def\nat{Nature}%
\def\physrep{Phys.~Rep.}%
\def\head{ .ps \vbox to 0pt{\vss \hbox to 0pt{\hskip 440pt\rm
      LA-UR-10-07069\hss} \vskip 25pt}} 

\def \spose#1{\hbox  to 0pt{#1\hss}}  
\def \lta{\mathrel{\spose{\lower 3pt\hbox{$\sim$}}\raise 2.0pt\hbox{$<$}}}
\def \gta{\mathrel{\spose{\lower 3pt\hbox{$\sim$}}\raise 2.0pt\hbox{$>$}}}

\title[Cosmic rays in cosmological simulations]
{The effects of cosmic rays on the formation of Milky Way-mass galaxies in a cosmological context}

\author[T. Buck \etal] {Tobias Buck$^{1}$\thanks{E-mail:
    tbuck@aip.de},
    Christoph Pfrommer$^{1}$,
    R\"udiger Pakmor$^{2}$,
    Robert J. J. Grand$^{2}$,
    \newauthor{\& Volker Springel$^{2}$}\\
$^1$Leibniz-Institut f\"ur Astrophysik Potsdam (AIP), An der Sternwarte 16, D-14482 Potsdam, Germany\\
$^2$Max-Planck-Institut f\"ur Astrophysik, Karl-Schwarzschild-Str. 1, D-85748, Garching, Germany
}

\begin{document}

\date{Accepted 2020 June 30. Received 2020 June 30; in original form 2019 October 29}

\pagerange{\pageref{firstpage}--\pageref{lastpage}} \pubyear{2019}

\maketitle

\label{firstpage}


\begin{abstract}
  We investigate the impact of cosmic rays (CR) and different modes of CR
  transport on the properties of Milky Way-mass galaxies in cosmological
  magneto-hydrodynamical simulations in the context of the AURIGA project. We
  systematically study how advection, anisotropic diffusion and additional
  Alfv\'en-wave cooling affect the galactic disc and the circum-galactic medium
  (CGM). Global properties such as stellar mass and star formation rate vary
  little between simulations with and without various CR transport physics,
  whereas structural properties such as disc sizes, CGM densities or
  temperatures can be strongly affected. In our simulations, CRs affect the
  accretion of gas onto galaxies by modifying the CGM flow structure. This
  alters the angular momentum distribution which manifests itself as a
  difference in stellar and gaseous disc size. The strength of this effect
  depends on the CR transport model: CR advection results in the most compact
  discs while the Alfv\'en-wave model resembles more the AURIGA model. The
  advection and diffusion models exhibit large ($r\sim50$ kpc) CR
  pressure-dominated gas haloes causing a smoother and partly cooler CGM. The
  additional CR pressure smoothes small-scale density peaks and compensates for
  the missing thermal pressure support at lower CGM temperatures. In contrast,
  the Alfv\'en-wave model is only CR pressure dominated at the disc-halo
  interface and only in this model the gamma-ray emission from hadronic
  interactions agrees with observations. In contrast to previous findings, we
  conclude that details of CR transport are critical for accurately predicting
  the impact of CR feedback on galaxy formation.
\end{abstract}

\noindent
\begin{keywords}

MHD - cosmic rays - galaxies: formation - galaxies: evolution - galaxies: structure - methods: numerical

 \end{keywords}

\section{Introduction} \label{sec:intro}

The formation of galaxies is a multi-scale, multi-physics problem and understanding the details of the physical processes involved is one of the most challenging problems in theoretical astrophysics. Cosmological simulations and semi-analytic studies have demonstrated that feedback from stellar winds and radiation fields, supernovae, and active galactic nuclei (AGNs) are key processes in shaping the structure of galaxies \citep[e.g.][]{Brook2012,Stinson2013,Puchwein2013,Marinacci2014,Vogelsberger2014,Henriques2015,Schaye2015,Dubois2016,Kaviraj2017,Pillepich2018,Hopkins2018}. These processes effectively drive galactic winds, move gas and metals out of galaxies into the intergalactic medium, regulate the star formation rate (SFR) down to the observed low rates or completely quench it in elliptical galaxies, and balance radiative cooling in the centers of galaxy clusters \citep{Kravtsov2012,Battaglia2012,Battaglia2012b,Battaglia2013,McCarthy2014,McCarthy2017,Dolag2016,Weinberger2017}.

While the latest galaxy formation models are quite successful in reproducing key observables of realistic galaxies \citep[e.g.][]{Wang2015,Sawala2016,Grand2017,Hopkins2018,Buck2019b}, most feedback prescriptions are modelled empirically, calibrated against observed scaling relations which limits the predictive power of the corresponding calculations. In particular, resolution requirements of hydrodynamical simulations of galaxy formation made it necessary to implement feedback relatively coarsely: simulations base their feedback prescriptions on explicit sub-grid formulations which model the unresolved, multiphase structure of the interstellar medium (ISM) \citep{Springel2003,Schaye2008}.
The details of the driving mechanisms behind galactic winds and outflows are still unknown and implementations remain phenomenological \citep{Oppenheimer2006}. On larger scales, feedback from AGNs has been invoked in order to balance star formation in galaxy clusters. Here, accretion rates onto the black hole are estimated by the Bondi prescription and feedback energy is injected in form of pure thermal energy \citep{DiMatteo2005,Springel2005} or they involve chaotic cold accretion \citep{Gaspari2013} or AGN feedback might be modelled slightly more complex \citep{Weinberger2017,Dave2019}.

Another obvious source of galactic feedback might be due to the energy and momentum deposition of the ultraviolet radiation of the stars. Radiation pressure acting on dust grains and the atomic lines in dense gas has been argued to transfer enough momentum to the gas in order to exceed the escape velocity and drive winds \citep{Murray2005,Thompson2005}. However, direct radiation-hydrodynamical simulations in simplified set-ups \citep{Krumholz2012} or in isolated galaxy simulations \citep{Rosdahl2015} do not produce strong radiation pressure driven winds, suggesting that radiation feedback is less effective and more gentle than widely assumed \citep[but see also][]{Emerick2018}.

On the other hand, cosmological simulations often neglect feedback from relativistic particles, so-called cosmic rays (CRs), which provide another source of non-thermal feedback. Such a particle population can be created by diffusive shock acceleration at expanding supernova remnants \citep[e.g.][]{Blandford1987, Jubelgas2008} or in AGN-powered jets \citep[e.g.][]{Sijacki2008,Ehlert2018}. CRs and magnetic fields are observed to be in pressure equilibrium with the turbulence in the mid-plane of the Milky Way \citep{Boulares1990} and the pressure forces of the CRs might be able to accelerate the ISM and drive powerful galactic outflows as suggested by a number of theoretical works \citep{Ipavich1975,Breitschwerdt1991,Breitschwerdt2002,Zirakashvili1996,Ptuskin1997,Socrates2008,Everett2008,Samui2010,Dorfi2012} and local three-dimensional (3D) simulations of the ISM \citep{Hanasz2013,Girichidis2016,Simpson2016}. 

\begin{table*}
\label{tab:props}
\begin{center}
\caption{Properties of the main galaxies: \normalfont{We state the virial mass, $M_{200}$, virial radius, $R_{200}$, the stellar mass, $M_{\rm star}$, and the gas mass, $M_{\rm gas}$, as well as the disc and bulge masses, $M_{\rm d}$, $M_{\rm b}$, as resulting from a combined exponential  plus \citet{Sersic1963} fit to the azimuthally averaged surface density profile. We further report the disc scale length, $R_{\rm d}$, the bulge effective radius, $R_{\rm eff}$, the bulge sersic index, $n$, and the disc-to-total ratio D/T of that fit.}}
\begin{tabular}{l c c c c c c c c c c}
		\hline\hline
		simulation & $M_{200}$ & $R_{200}$ & $M_{\rm star}$ & $M_{\rm gas}$ & $M_{\rm d}$ & $R_{\rm d}$ & $M_{\rm b}$  & $R_{\rm eff}$ & $n$ & D/T \\
		  & [$10^{12}\Msun$] & [kpc] & [$10^{10}\Msun$] & [$10^{10}\Msun$] & [$10^{10}\Msun$] & [kpc]  & [$10^{10}\Msun$]& [kpc] & & \\
		\hline
		\multicolumn{11}{c}{Auriga-6 (Au6)}\\
		\hline
		noCR & 1.02 & 212.39 & 4.36 & 7.88 & 4.17 & 4.53 & 0.41 & 0.89 & 0.69 & 0.91 \\
		CRdiffalfven & 1.06 & 215.18 & 5.54 & 9.12 & 4.37 & 2.84 & 1.03 & 0.82 & 0.83 & 0.81\\
		CRdiff & 1.07 & 215.43 & 5.81 & 8.91 & 1.03 & 4.37 & 4.64 & 1.14 & 1.11 & 0.18\\
		CRadv & 1.09 & 216.71 & 6.19 & 9.47 & 1.11 & 4.00 & 4.79 & 1.11 & 1.54 & 0.19\\
		\hline
		\multicolumn{11}{c}{Auriga-L8 (AuL8)}\\
		\hline
		noCR & 0.84 & 199.02 & 4.82 & 7.00 & 4.39 & 3.65 & 0.41 & 0.73 & 0.58 & 0.91\\
		CRdiffalfven & 0.83 & 197.96 & 4.72 & 6.06 & 4.08 & 3.76 & 0.67 & 0.90 & 0.91 & 0.86\\
		CRdiff & 0.83 & 198.48 & 4.51 & 6.65 & 0.00 & 0.00 & 4.51 & 4.33 & 1.71 & 0.00\\
		CRadv & 0.85 & 199.33 & 4.58 & 7.76 & 0.00 & 0.00 & 4.58 & 3.79 & 1.49 & 0.00\\
        \hline
\end{tabular}
\end{center}
\end{table*}

In comparison to other feedback mechanisms, CRs have a number of advantageous properties: 
(i) the CR pressure drops less quickly upon adiabatic expansion than the thermal pressure due to their softer equation of state ($P_{\rm CR} \propto \rho^{\gamma_{\rm CR}}$ with $\gamma_{\rm CR} = 4/3$),
(ii) CR cooling is generally less efficient than the radiative cooling of a thermal plasma \citep{Ensslin2007} and thus acts on longer time-scales compared to thermal energy,
(iii) the non-thermal energy of CRs is not detectable through thermal observables or X-ray emission, therefore the (temporary) storage of feedback energy in CRs also avoids problems with the overproduction of these observables
(iv) they can maintain the outflows in a warm/hot state because the resonant production of Alfv\'en waves through the streaming instability \citep{Kulsrud1969} and the dissipation of wave energy with various plasma physical wave damping processes energises galactic winds as they rise in the galactic haloes.

A number of works have implemented CRs into 3D hydrodynamic simulations of galaxy formation and have demonstrated the ability of CRs to drive winds and regulate star formation  \citep{Jubelgas2008,Uhlig2012,Booth2013,Salem2014,Pakmor2016,Ruszkowski2017,Pfrommer2017b,Jacob2018,Chan2019}. CRs do not couple to the thermal gas via particle-particle collisions but via particle-wave interactions as fast streaming CRs along the magnetic field resonantly excite Alfv\'en waves. CRs are then able to scatter off of these waves which isotropise their distribution function in the wave frame, thus transferring energy and momentum to the thermal plasma and exerting a pressure onto the gas. Thereby, CRs not only impart momentum to the ISM at the launching sites but continuously re-power winds via thermal and dynamic coupling of plasma and CRs.
   
As a result CRs might explain the observed low SFRs in giant elliptical galaxies located at the centers of galaxy groups and clusters. In the absence of any heating processes, the hot gaseous atmosphere of these objects is expected to efficiently cool and form stars at very high rates \citep[up to a few hundred \Msun yr$^{-1}$, e.g.][]{Peterson2006}. However, observed SFRs are much below these expected rates which is why AGN feedback has been invoked to balance radiative cooling. While theoretical considerations have shown that AGN feedback energies are sufficient, the exact coupling mechanism is still under debate \citep{McNamara2007}. While several physical processes have been proposed to mediate the heating \citep[amongst others the dissipation of turbulent energy powered by the AGN,][]{Zhuravleva2014} a promising alternative is given by CRs. A net outward flux of streaming CRs can resonantly excite Alfv\'en waves that experience non-linear Landau damping or decay via a cascading process as a result of strong external turbulence and eventually dissipate locally \citep{Loewenstein1991,Guo2008,Ensslin2011,Pfrommer2013,Wiener2013,Jacob2017,Jacob2017b,Ruszkowski2017b,Ehlert2018}.

By now there are several (magneto-)hydrodynamics (MHD) simulation codes capable of solving the details of the CR proton acceleration and transport in galaxies
and galaxy clusters: 
the Eulerian mesh codes \textsc{cosmocr} \citep{Miniati2001}, \textsc{zeus-3d} \citep{Hanasz2003}, the smoothed particle hydrodynamics code \textsc{gadget-2}
\citep{Pfrommer2006,Ensslin2007,Jubelgas2008}, the adaptive mesh refinement codes \textsc{ramses} \citep{Booth2013, Dubois2016}, \textsc{enzo} \citep{Salem2014}, \textsc{flash} \citep{Girichidis2016}, and \textsc{pluto} \citep{Mignone2018}, the moving-mesh code \textsc{arepo} \citep{Pakmor2016b,Pfrommer2017} and the mesh-free Lagrangian finite mass code \textsc{gizmo} \citep{Chan2019}.  Here, we use the \textsc{arepo} code \citep{Springel2010,Pakmor2016c} combined with the numerical implementations of CR physics \citep{Pfrommer2017,Pakmor2016b} to simulate the formation of Milky Way (MW) like galaxies in a cosmological context. 

This paper is organized as follows: In \Sec{sec:sim} we describe the simulation setup and the different implementations of CR treatment. In \Sec{sec:props} we study the central stellar and gaseous discs focussing on the differences and similarities in properties across various CR physics variants. We further investigate here the accretion of gas onto the main galaxy and the successive build-up of angular momentum. In \Sec{sec:CGMprops} we turn to analyse the effects of CRs on the properties and structure of the CGM. We finish our analysis in \Sec{sec:obs} by comparing a direct observable, namely the gamma-ray luminosity of the simulated galaxies, to observations. In \Sec{sec:dis} we conclude this paper with a discussion, compare our results to previous work, and summarize our results in \Sec{sec:conc}.

\section{Cosmological Simulations} \label{sec:sim}

For this work we simulate the formation of two MW-like disc galaxies from cosmological initial conditions taken from the AURIGA project \citep{Grand2017,Grand2019}. Simulations are performed with the second-order accurate, adaptive moving-mesh code AREPO \citep{Springel2010,Pakmor2016c} which includes important physical prescriptions to model galaxy formation in a cosmological setup. For completeness we describe the most important models below but refer the reader to the works of \citet{Grand2017}, \citet{Pakmor2016b} for more technical details on the AURIGA galaxy formation model and to \citet{Pfrommer2017} for details on the CR physics.

The AURIGA model includes primordial and metal-line cooling with self-shielding corrections and the spatially uniform UV background model of \citet{Faucher2009} is included \citep[for more details see][]{Vogelsberger2013}. The interstellar medium (ISM) is modelled with an effective equation of state \citep{Springel2003} and star-forming gas is treated as a two phase medium. Star formation occurs in thermally unstable gas for densities higher than a threshold density of $n_{\rm th} = 0.13\mbox{ cm}^{-3}$ in a stochastic manner where the probability scales exponentially with time in units of the star formation timescale ($t_{\rm sf}=2.2$ Gyr in the AURIGA model).

\begin{figure*}
\vspace{-.5 cm}
\begin{center}
\includegraphics[width=\textwidth]{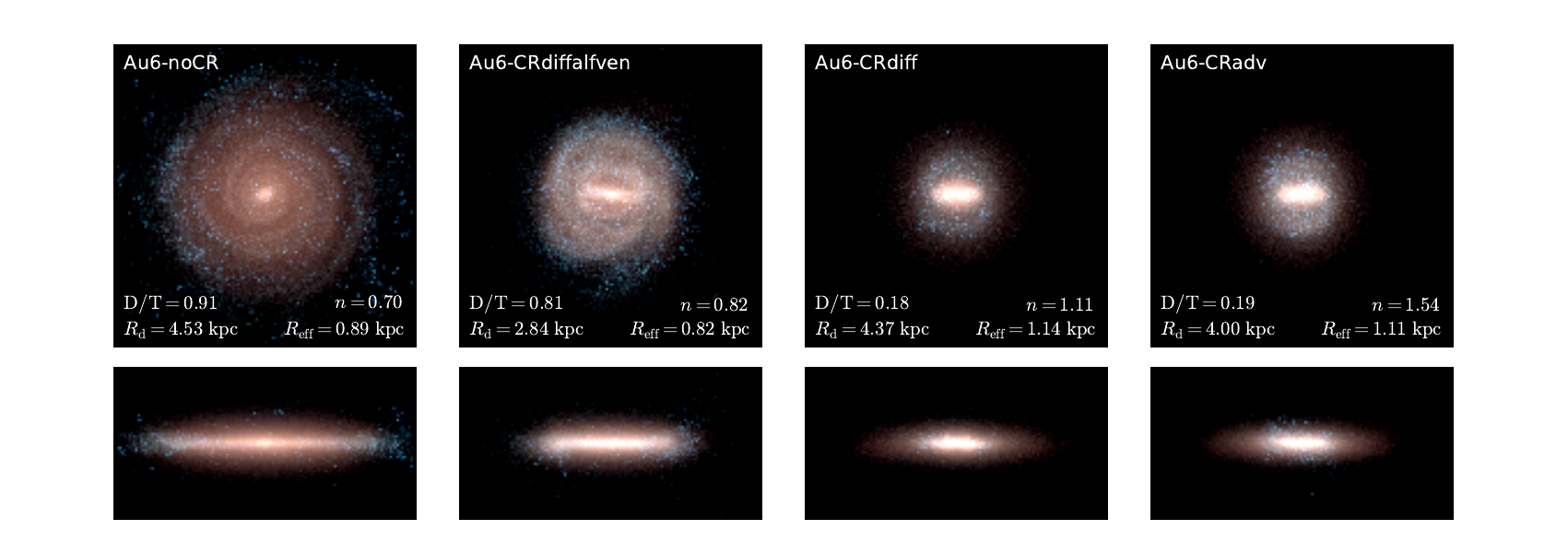}
\includegraphics[width=\textwidth]{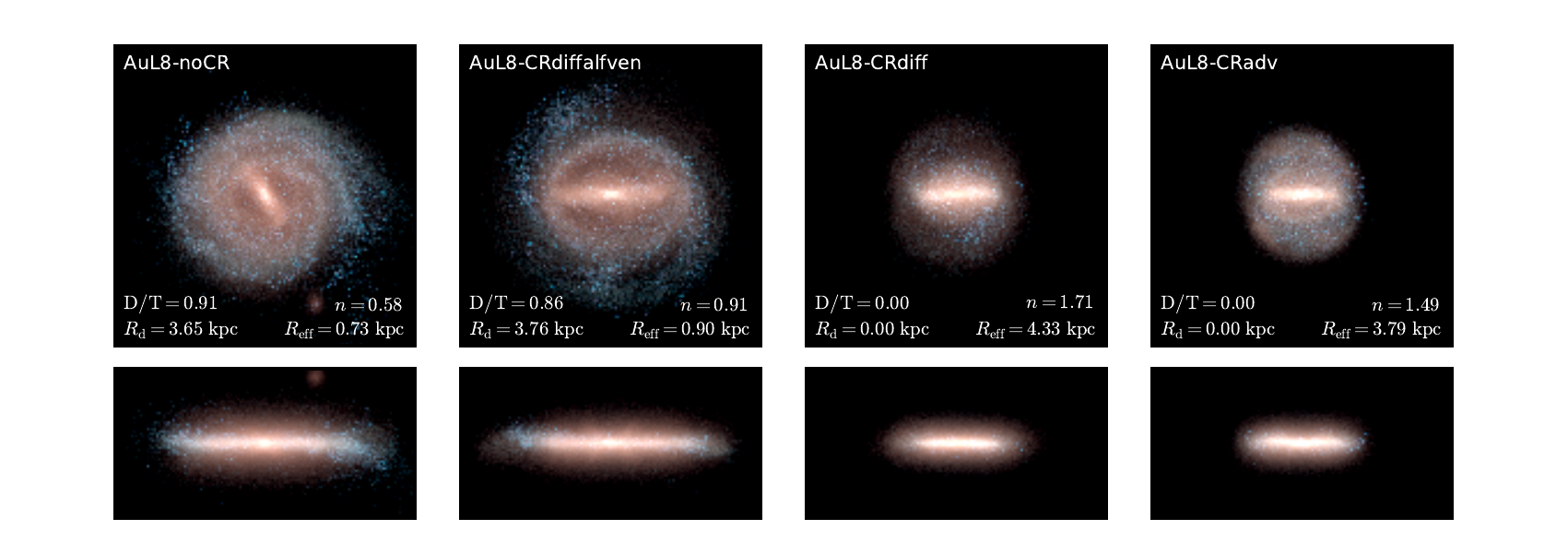}
\end{center}
\vspace{-.35cm}
\caption{Face-on and edge-on projected stellar density at $z=0$ for the eight simulations. The upper panel shows the galaxy Au6 and the lower panel shows AuL8. From left to right we show the four different variants of physics: (i) the fiducial AURIGA simulations without CRs, (ii) the simulations with CR advection and anisotropic CR diffusion and Alfv\'en cooling/heating enabled, (iii) the CR anisotropic diffusion and CR advection simulations, and (iv) the pure CR advection simulations. The images are synthesized from a projection of the $K$-, $B$- and $U$-band luminosity of stars, which are shown by the red, green and blue colour channels, in logarithmic intervals, respectively. Younger (older) star particles are therefore represented by bluer (redder) colours. In each face-on panel we note structural properties of the stellar disc resulting from a surface density fit of a combined exponential plus \citet{Sersic1963} profile (see Fig.\ \ref{fig:surf_den_fit}). The plot dimensions are 50 kpc $\times$ 50 kpc and 50 kpc $\times$ 25 kpc, respectively.}
\label{fig:rgb}
\end{figure*}

Each star particle in this model represents a single stellar population (SSP) characterised by age and metallicity assuming a \citet{Chabrier2003} initial mass function (IMF). The mass loss and metal return of each star particle is calculated for supernovae of type II (SNII), SNIa and AGB stars at each time-step. The mass and metal yields are taken from \citet{Karakas2010} for AGB stars, and \citet{Portinari1998} for core collapse SNe. Mass and metals are distributed among nearby gas cells with a top-hat kernel. SNIa events are calculated using a delay time distribution function and their mass and metal return follows the yield tables from \citet{Thielemann2003} and \citet{Travaglio2004}. The number of SNII is calculated from the number of stars in the mass range $8-100\Msun$ and feedback is assumed to occur instantaneously. An active gas cell (star-forming) is probabilistically chosen to either form a star particle or become a site for SNII events \citep[see][]{Vogelsberger2013} in which case a wind particle is launched in an isotropically random direction. The wind velocity is coupled to the one-dimensional velocity dispersion of the dark matter halo. The wind particle interacts only gravitationally until either a gas cell with a density below $0.05$ times the threshold density for star formation is reached, or the maximum travel time is exceeded. Meeting either of the criteria, the wind particle re-couples and deposits its mass, metals, momentum and thermal energy into the gas cell in which it is located \citep[see][and references therein for more details]{Grand2017}. Note, CRs are directly injected into neighbouring cells of a supernova and as such wind particles neither transport CR energy nor the magnetic field.

Active galactic nuclei feedback is implemented following \citet{Springel2005}. The mass growth from gas accretion is described by Eddington-limited Bondi-Hoyle-Lyttleton accretion \citep{Bondi1944,Bondi1952} in addition to a term that models the radio mode accretion \citep[see][for more details]{Grand2017}. Magnetic fields are treated with ideal MHD \citep{Pakmor2011,Pakmor2013} and are seeded at redshift $z=127$ with a homogeneous magnetic field of $10^{-14}$ (comoving) Gauss. The divergence cleaning scheme of \citet{Powell1999} is implemented to ensure that the divergence of the magnetic field vanishes.

Finally, CRs are modelled as a relativistic fluid with a constant adiabatic index of $4/3$ in a two-fluid approximation \citep{Pfrommer2017}. CRs are generated at core-collapse supernovae remnants by instantaneously injecting all CR energy produced by the star particle into its surroundings immediately after birth. The energy efficiency of the injection is set to $\zeta_{\rm SN} = 0.1$. Following \citet{Pfrommer2017}, we assume an equilibrium momentum distribution for the CRs to model their cooling via Coulomb and hadronic interaction with the ambient gas. 

To bracket the uncertainties of CR transport, we simulate three different models with different variants of CR physics \citep[similar to][]{Wiener2017} and one model without CRs. To explain the differences of our CR models, we briefly review the main aspects of CR hydrodynamics. While individual CRs move close to the speed of light, frequent resonant CR interactions with Alfv\'en waves causes their distribution function to (nearly) isotropise in the frame of the Alfv\'en waves such that the CR energy is transported as a superposition of CR advection with the gas, anisotropic streaming with Alfv\'en waves along the magnetic field and diffusion with respect to the wave frame so that the time evolution equation of the CR energy density $\eps_\CR$ in the one-moment formulation of CR transport reads as follows:
\begin{align}
  \frac{\partial \eps_\CR}{\partial t}
  &+ \bnabla\bcdot
  \big[
    \underbrace{\eps_\CR \bvel}_{\rmn{advection}}
    + \underbrace{(\eps_\CR+P_\CR)\bvel_{\rmn{st}}}_{\rmn{streaming}}
    - \underbrace{\kappa_\eps \bb \left( \bb \bcdot \bnabla \eps_\CR\right)}_{\rmn{anisotropic\ diffusion}}\big]
  \nonumber\\
    &= - \underbrace{P_\CR\,\bnabla \bcdot \bvel}_{\rmn{adiab.\ changes}}
      - ~\underbrace{\left|\bvel_{\rmn{A}} \bcdot \bnabla P_\CR\right|}_{\rmn{Alfven\ cooling}}
      + \underbrace{\Lambda_\CR + \Gamma_\CR}_{\rmn{losses\ \&\ sources}}.
      \label{eq:ecr}
\end{align}
Here, $\bvel$ denotes the gas velocity, $\bvel_{\rmn{A}}=\bB/\sqrt{4\upi\rho}$ is the Alfv\'en velocity, $\bB$ is the magnetic field, $\rho$ is the gas mass density, $\bvel_{\rmn{st}}$ is the CR streaming velocity, 
\begin{eqnarray}
  \label{eq:vstream}
  \bvel_{\rmn{st}} = -\bvel_{\rmn{A}}\, \rmn{sgn}(\bB\bcdot\bnabla P_\CR)
  = -\frac{\bB}{\sqrt{4\upi\rho}}\,
  \frac{\bB\bcdot\bnabla P_\CR}{\left|\bB\bcdot\bnabla P_\CR\right|},
\end{eqnarray}
implying that the CR streaming velocity is oriented along magnetic fields lines down the CR pressure gradient with a velocity that corresponds in magnitude to $\bvel_{\rmn{A}}$,$P_\CR$ is the CR pressure, $\kappa_\eps$ is the kinetic energy-weighted spatial CR diffusion coefficient, $\bb=\bB/|\bB|$ is the unit vector along the local magnetic field, and $\Lambda_\CR$ and $\Gamma_\CR$ are non-adiabatic CR losses and sources.

Note that CR streaming and diffusion are both anisotropic transport processes along the mean magnetic field and oriented down the CR gradient. While the streaming term advects CRs with the frame of Alfv\'en waves and maintains CR gradients, diffusion is a dispersive process (owing to the second gradient in the bracket of Eq.~\eqref{eq:ecr}) so that the CR gradient weakens over time, implying that the streaming and diffusion fluxes cannot be the same at all times \citep{Wiener2017}. Most importantly, CR diffusion exactly conserves the CR energy while CR streaming drains CR energy at a rate $\left|\bvel_{\rmn{A}} \bcdot \bnabla P_\CR\right|$ due to the excitation of resonant Alfv\'en waves.

\begin{figure*}
\vspace{-.45 cm}
\begin{center}
\includegraphics[width=.33\textwidth]{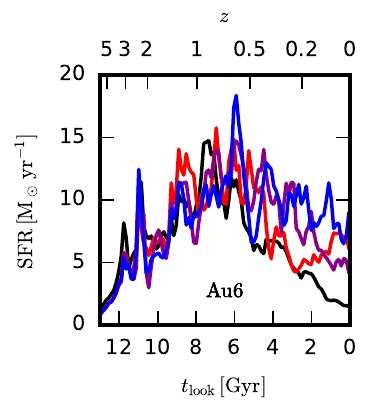}
\includegraphics[width=.33\textwidth]{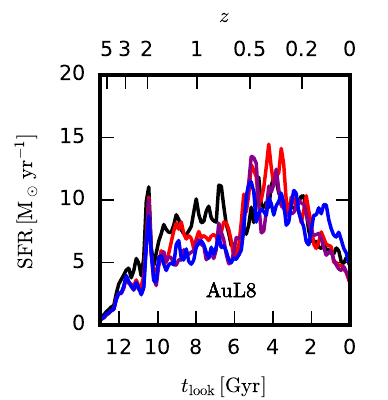}
\includegraphics[width=.33\textwidth]{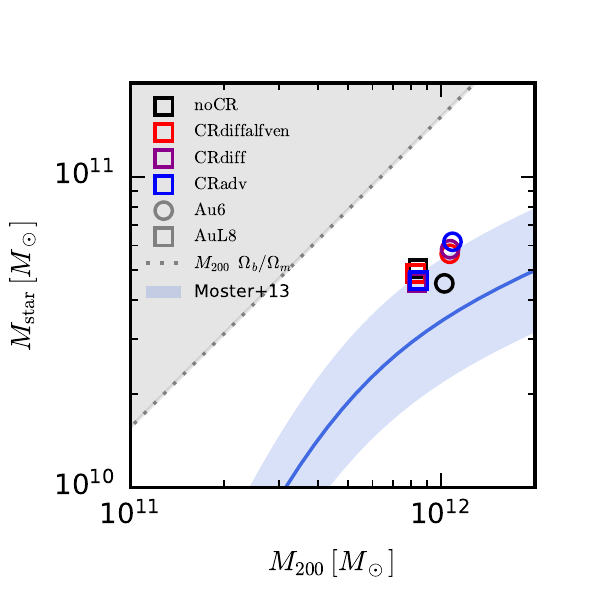}
\end{center}
\vspace{-.35cm}
\caption{Star formation histories (SFH) and the stellar mass-halo mass relation (right panel) for the AURIGA haloes. We show the SFHs for galaxy Au6 (left panel) and for AuL8 (middle panel), respectively. In the right panel, we show the final stellar mass of our simulations at $z=0$ vs their final halo mass. In each panel we compare the SFH/stellar masses of the four different physics variants, the fiducial Auriga model shown in black, the cosmic ray advection run in blue, the run with additional anisotropic CR diffusion in purple and the run which additionally accounts for CR Alfv\'en wave cooling in red. In the right panel, the light blue line shows the abundance matching result taken from \citet{Moster2013} while the gray dotted line shows the cosmic baryon fraction.}
\label{fig:sfr}
\end{figure*}

In all of our models we omit the CR streaming term on the left-hand side of Eqn.~\eqref{eq:ecr}, which can be most accurately solved with the two-moment method of CR transport \citep{Jiang2018,Thomas2019}. In all three CR models, we account for CR advection and adiabatic changes of the CR energy. Our models are defined as follows:
\begin{itemize}
\item[(i)] \textbf{noCR}: fiducial AURIGA galaxy formation model without CRs.
\item[(ii)] \textbf{CRadv}: CR advection model where CRs are only advected with the gas.
\item[(iii)] \textbf{CRdiff}: CR diffusion model where CRs are advected with the gas but are further allowed to anisotropically diffuse relative to the rest frame of the gas with a diffusion coefficient of $\kappa_\parallel = 10^{28}$~cm$^2$~s$^{-1}$ along the magnetic field and no diffusion perpendicular to it \citep{Pakmor2016b}.
\item[(iv)] \textbf{CRdiffalfven}: anisotropic CR diffusion model with the additional inclusion of the Alfv\'en-wave cooling term that arises due to the energy transfer from CRs to Alfv\'en waves that are self-excited through the resonant CR streaming instability \citep{Kulsrud1969}.
\end{itemize}
While CR diffusion is thought to describe the transport of high-energy CRs above $\sim200$ GeV, at lower energies the transport transitions to mainly CR streaming with self-generated Alfv\'en waves \citep{Evoli2018}, although the role of scattering in external turbulence is not yet settled \citep{Zweibel2017}. While the process of CR diffusion conserves CR energy, additionally accounting for the Alfv\'en-wave cooling term emulates and approximates CR streaming. While this approximation is justified in cases for which the diffusion and streaming fluxes match each other, solutions will necessarily deviate if this condition is not fulfilled \citep{Wiener2017}. Future work is needed to clarify how the explicit inclusion of CR streaming in the presence of different wave damping processes changes the picture presented in this work.

\begin{figure*}
\vspace*{-.4cm}
\begin{center}
\raggedleft
\includegraphics[width=.925\textwidth]{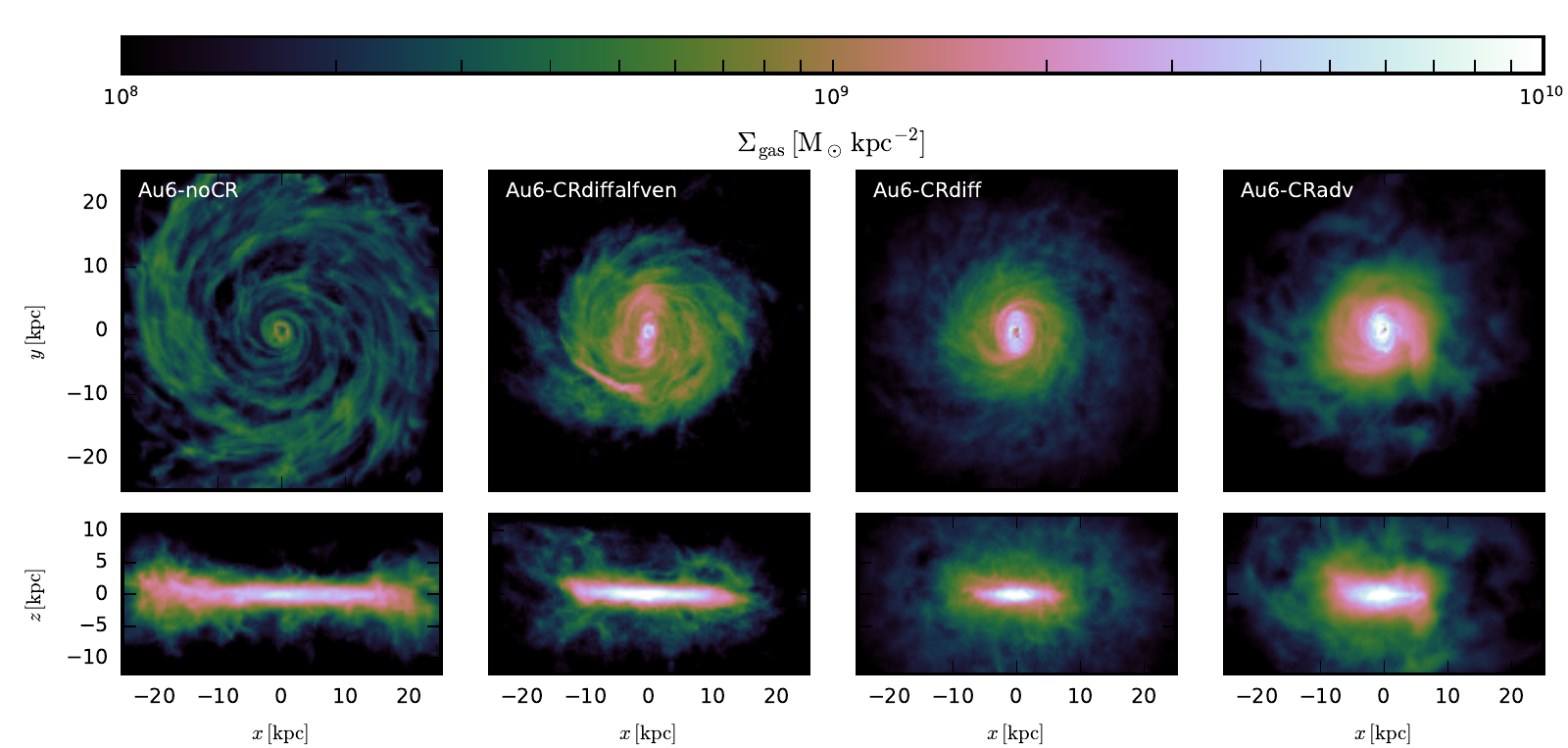}
\includegraphics[width=.925\textwidth]{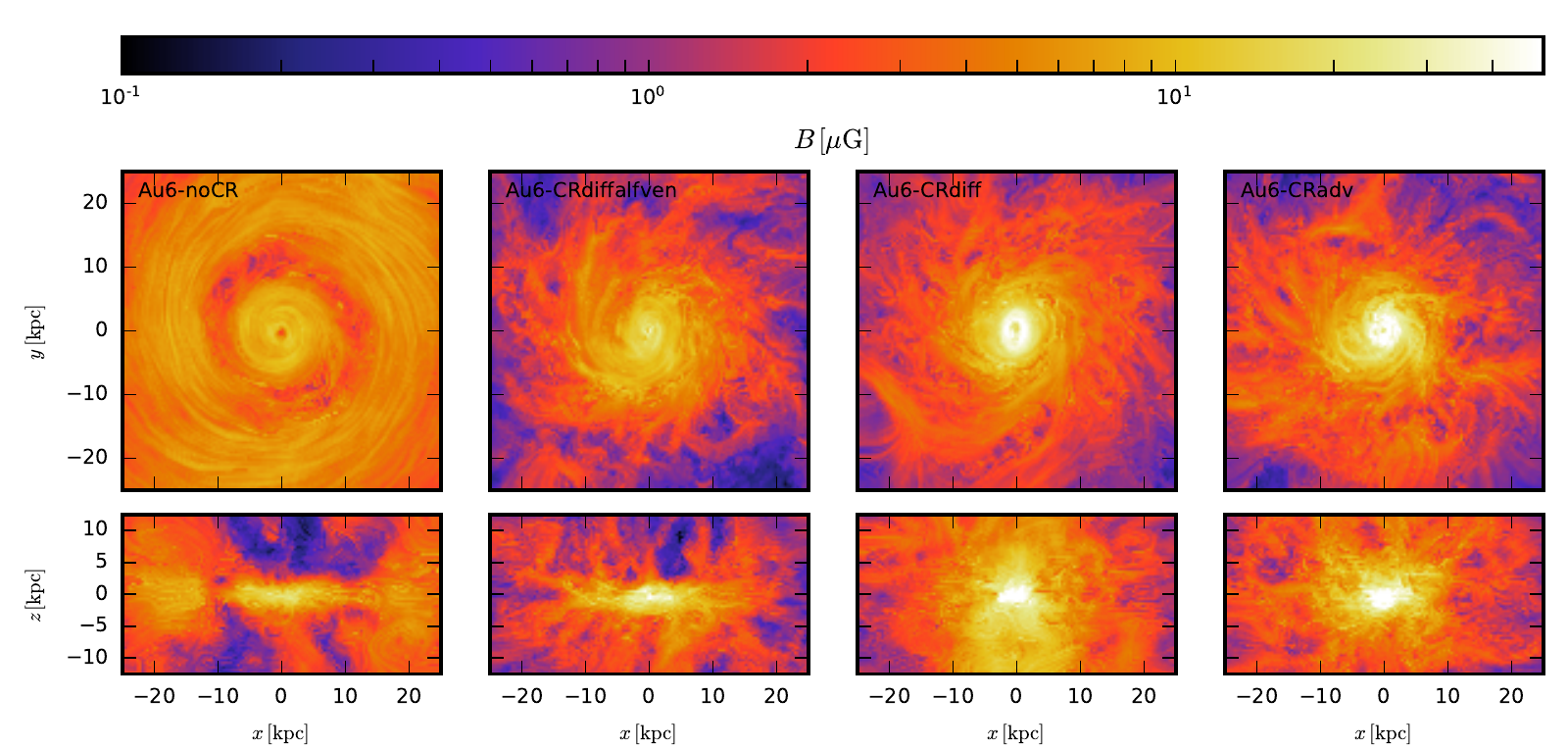}
\includegraphics[width=.71\textwidth]{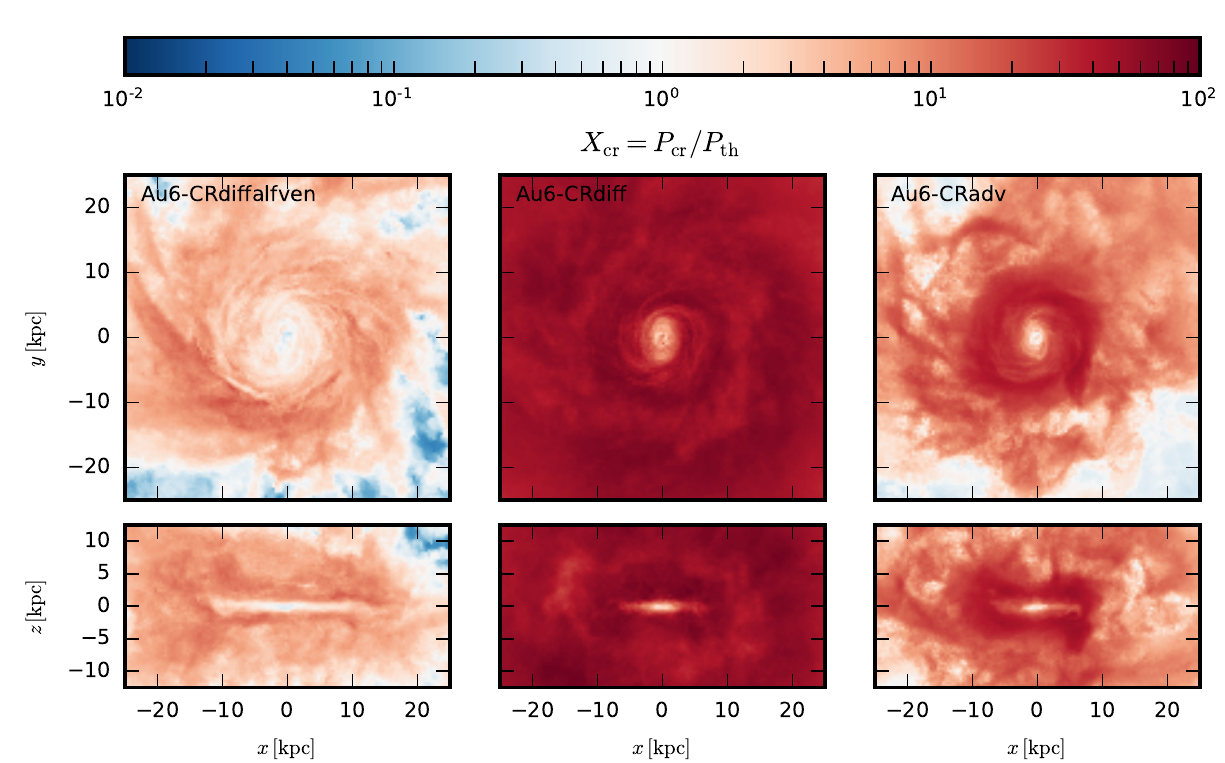}
\end{center}
\vspace{-.35cm}
\caption{From top to bottom we show the gas surface density, the volume-weighted magnetic field strength, $\sqrt{\langle\bB^2\rangle_V}$, and the CR-to-thermal pressure ratio, $X_{\rm cr} = \left<P_{\rm cr}\right>_V / \left<P_{\rm th}\right>_V$, of the different physics variants (left to right) in the Au6 simulation in face-on and edge-on projections. The projection depth is $25$ kpc.}
\label{fig:gas}
\end{figure*}

\begin{figure*}
\vspace{-.25 cm}
\begin{center}
\includegraphics[width=1.0\textwidth]{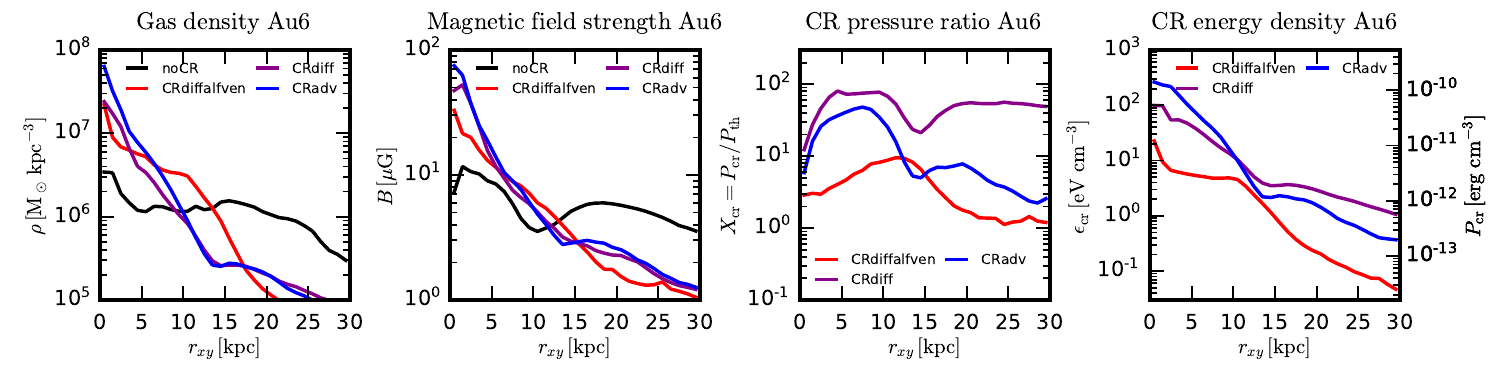}
\includegraphics[width=1.0\textwidth]{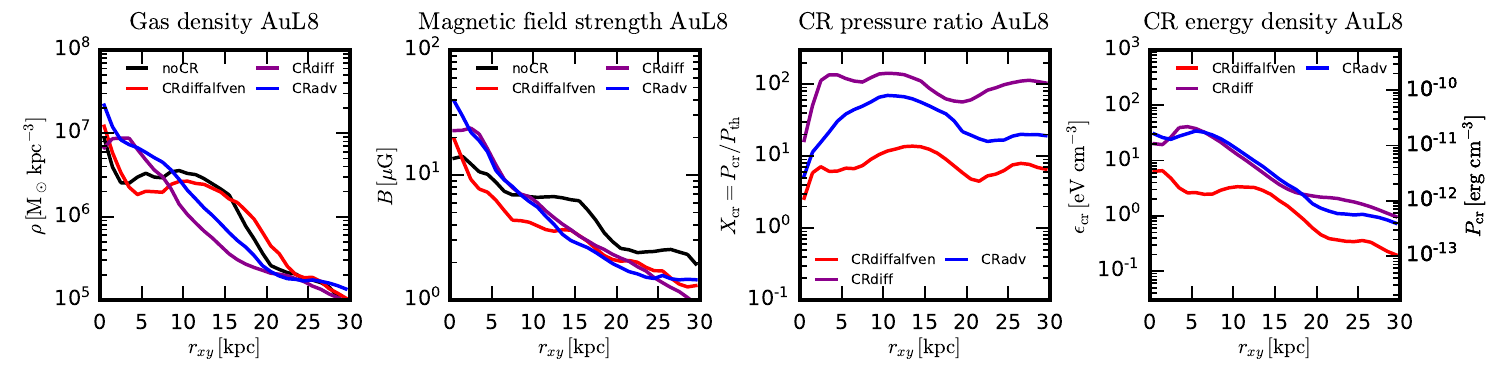}
\end{center}
\vspace{-.6cm}
\caption{Radial profiles of gas density, magnetic field strength, CR-to-thermal pressure ratio, and CR pressure and energy density (left to right) in cylindrical shells of height $\vert z\vert=3$~kpc and width $\Delta r_{xy}=1$ kpc. The upper panels show results for the model galaxy Au6 and lower panels for AuL8. Different physics variants are shown with differently colored lines. The fiducial AURIGA run is shown with a black line, the Alfv\'en run in red, the CR diffusion run in magenta and the CR advection run in blue.}
\label{fig:prof}
\end{figure*}

\section{Galaxy disc properties} \label{sec:props}

\subsection{Stellar disc}

In Fig.\ \ref{fig:rgb} we present face-on and edge-on projections of all eight simulations at $z = 0$ where the upper panels show the four different simulations of the Au6 halo and lower panels of AuL8. The images are a composition of the K-, B- and U-band luminosities (mapped to the red, green and blue colour channels), which indicate the distribution of younger (bluer colours) and older (redder colours) star particles, respectively. All simulations reveal a star-forming disc component with additional clear non-axisymmetric structures such as bars and spiral arms. For the fiducial noCR and the CRdiffalfven model the stellar disc is radially extended and thin. In contrast to that, the CRdiff and CRadv models result in more compact stellar discs further indicated by the lower ratio of D/T between the stellar disc mass (D) and the total stellar mass of the galaxy (T) shown in the lower right corner of each panel. Nevertheless, from the edge-on view these simulations are still identifiable as disc galaxies.

Despite the obvious differences in morphology, the total stellar mass in each model is almost the same as shown by Fig.\ \ref{fig:sfr}. Here, we investigate the star formation history (SFH) and the stellar mass-halo mass relation of the galaxies in the four different models. The SFH of halo Au6 is shown in the left most panel, the one of halo AuL8 in the middle panel and the right panel shows the stellar mass as function of the halo mass in comparison to abundance matching results from \citet{Moster2013}. Within their intrinsic scatter, the SFHs of each physics run do not differ much and the total stellar mass of each galaxy varies only within a factor of $\lesssim1.5$ at the end of each run (see also Table \ref{tab:props}). For halo AuL8 we find that the CRs slightly suppress the stellar mass growth already at early times ($z\sim2$) as can be appreciated from the suppressed SFR (middle panel). At redshift zero the CRdiffalfven run has very similar stellar mass compared to the AURIGA run caused by a late time ($z\lesssim0.2$) burst in star formation (see red line in the middle panel) whereas the CRdiff and CRadv run have slightly lower stellar mass. Halo Au6 on the other hand acquires slightly more stellar mass in the CR runs compared to the fiducial AURIGA model, at late times ($z<0.5$) also visible in the enhanced late time SFR (left panel of Fig.\ \ref{fig:sfr}) for the CR runs. 

We further examined the total amount of gas as well as the cold gas mass and found that both quantities do not change much across the different physics variants. We conclude that in our simulations structural disc properties can be significantly changed by CRs, global stellar properties, however, are robust across different CR physics variants and are not much affected by CRs.

\subsection{Gas disc}
\label{subsec:gasdisc}
The differences in stellar morphology are mainly a result of differences in the gaseous properties of the central galaxies. Figure \ref{fig:gas} shows face-on and edge-on projections of the gas surface density (upper panels), the magnetic field strength (middle panels) and the ratio of CR-to-thermal pressure, $X_{\rm cr}$, (lower panels) of the central gaseous disc of Au6. From left to right we show the noCR, the CRdiffalfven, the CRdiff and the CRadv run. The magnetic field is calculated as $B=\sqrt{\left<B^2\right>_V}$ and the value of $X_{\rm cr}$ is calculated as the ratio of volume-weighted pressure terms, $X_{\rm cr} = \left<P_{\rm cr}\right>_V / \left<P_{\rm th}\right>_V$. Figure \ref{fig:prof} complements this by showing the radial profiles in cylindrical bins of radial width $1$ kpc and height $\pm3$ kpc centered on the disc mid-plane for both galaxies Au6 (upper panels) and AuL8 (lower panels). The profiles have been obtained by adding up thermodynamic extensive quantities. This means, that we have volume-weighted \mbox{(energy-)}densities within a given concentric shell, while temperature averages are calculated via mass weighting.

Comparing the surface density maps of the four different physics variants we find that the CR runs show a more centrally concentrated, thicker gas disc with slightly higher surface densities within the disc region. This is further highlighted by the larger central densities in the left panels of Fig.\ \ref{fig:prof}. Furthermore, from Fig. \ref{fig:gas} we see how the CR pressure smoothes out density features in the disc, particularly in the CRadv and CRdiff runs; the CRdiffalfven run most closely resembles the fiducial AURIGA run. The CRadv run exhibits the most compact, thick and smooth gas disc where the additional CR pressure stabilizes and smoothes the gas. The lower panels of Fig.\ \ref{fig:gas} and the third panels from the left of Fig.\ \ref{fig:prof} show that in all three CR models the thermal pressure in the gas disc is sub-dominant in comparison to the CR pressure. This effect is most prominent in the CRadv and CRdiff runs where CRs can only cool adiabatically via Coulomb and hadronic interactions. This results in CR-to-thermal pressure ratios of $P_{\rm cr}/P_{\rm th}=X_{\rm cr}\gtrsim10$ in the disc region.

We observe slightly higher CR-to-thermal pressure ratios in the CRdiff compared to the CRadv run. At first, this result might be surprising because the CRs in the diffusion run are able to diffuse out of the disc into the halo. The reason for this is as follows: if $\vert \bnabla (P_{\rm cr} + P_{\rm th}) \vert > \vert \rho \bnabla \Phi \vert$ (where $P_{\rm th}$ is the thermal pressure and $\Phi$ is the gravitational potential), then the composite of CRs and thermal gas adiabatically expands and as a result the CR pressure will exceed the thermal pressure because of its softer equation of state; the CR pressure decreases at a slower rate in comparison to the thermal pressure. As the CRs diffuse above and below the galaxy midplane, they push gas out of the disc (via their gradient pressure force), thus lowering the gas density in the disc (see left panels of Fig.~\ref{fig:vert_prof}). Because the temperature in the star forming regions is set by the effective equation of state, the thermal pressure in the CRdiff model is lower (see right most panels of Fig.~\ref{fig:vert_prof}), and hence the ratio of $P_{\rm cr}/P_{\rm th}=X_{\rm cr}$ is larger in this model.

In the CRdiffalfven run on the other hand, the CRs are allowed to diffuse and to cool via the Alfv\'en wave cooling mechanism and thus their stabilizing pressure is less dominant compared to the CRadv run. Here we find typical values of $X_{\rm cr}$ ranging from three to ten within the central $5-10$ kpc of the disc (see Fig.\ \ref{fig:prof}). This allows for a shallower radial density profile within the disc that is then able to grow larger stellar discs. The right-most panel of Fig.\ \ref{fig:prof} shows the radial profiles of the disc CR pressure and the energy density, respectively, averaged over a height of $|z|=3$~kpc. The CR energy density is decreasing with radius and follows in general the shape of the gas density profile with breaks at $r_{xy}\sim14$~kpc and 20~kpc for Au6 and AuL8, respectively. At the solar radius ($\sim8$ kpc) the value of the CR energy density is $\epsilon_{\rm cr}\sim4-5$~eV~cm$^{-3}$ for Au6 ($\epsilon_{\rm cr}\sim2$~eV~cm$^{-3}$ for AuL8) in the CRdiffalfven model while the other two models show a factor of $\gsim3$ higher values. A more detailed comparison to observations follows in Section~\ref{sec:obs_comp}.
We caution that some of the drastic differences of the density profiles between the different variants of CR physics may be due to cosmic variance and the different accretion histories. In particular the differing density profiles in Au6 (Fig.\ \ref{fig:prof}) are reduced in AuL8 where the CRdiffalfven run's density profile follows much more closely the fiducial AURIGA run.

The middle panel of Fig.\ \ref{fig:gas} shows the magnetic field strength which looks very similar in the disc for all CR runs but varies drastically from the fiducial AURIGA runs which show a much smoother, ordered magnetic field. Looking at the second panels of Fig.\ \ref{fig:prof} we find that the magnetic field strength at the edge of the stellar disc ($\sim20$ kpc) is around 1.5 to 2.5 $\mu$G (except for the fiducial Au6 run which shows values of $\sim6$ $\mu$G). The magnetic field strength in our models increases towards the central regions of the disc. In the inner disc (at radii of $\lesssim5$ kpc) we find magnetic field strengths of $\gtrsim7-15~\mu$G ($\gtrsim9-20~\mu$G for Au6) depending on the CR model at hand.
These values of the magnetic field strength are in good agreement with estimates for the MW \citep{Haverkorn2006,Haverkorn2015,Sun2012,Pakmor2018} and local disc galaxies \citep{Beck2019}. In detail, the CR runs show a more structured magnetic field which follows closely the structure of the gas disc because the CRs act as a local feedback source while in the noCR run the wind feedback is non local. Thus, CRs are able to inject turbulence in the gas disc, imprinting more small scale structure onto the magnetic field. This feature is absent in the noCR runs and thus the magnetic field appears much more ordered. However, the halo magnetic field looks very different between the three CR runs. Interestingly, the vertical magnetic field extending into the halo in the CRdiff and CRadv is larger compared to the CRdiffalfven and AURIGA runs. Thus, we conclude that CR dynamics alters the dynamo process in comparison to pure MHD simulations and the higher density features in the CR runs lead to a more structured magnetic field in the gas disc (see also middle panels of Fig.\ \ref{fig:prof}). We note that different variants of CR transport seem not to affect the disc magnetic field much.

\begin{figure}
\vspace*{-.5cm}
\hspace*{-.5cm}
\includegraphics[width=1.25\columnwidth]{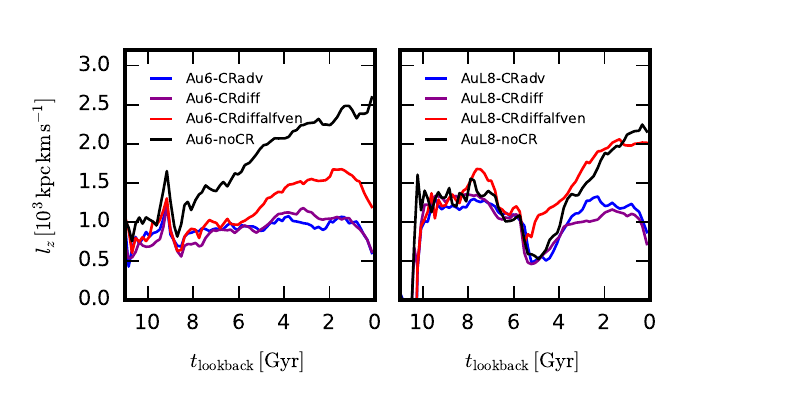}
\vspace{-.75cm}
\caption{Median specific angular momentum evolution of the gas that ends up in the stellar disc at redshift $z=0$. Au6 is shown in the left panel and AuL8 in the right. At lookback times of $t_{\rm lookback}\sim6$ Gyr AuL8 is undergoing a major merger causing the dip in the angular momentum evolution.}
\label{fig:ang_mom}
\end{figure}

\begin{figure*}
\vspace*{-.4cm}
\begin{center}
\includegraphics[width=\textwidth]{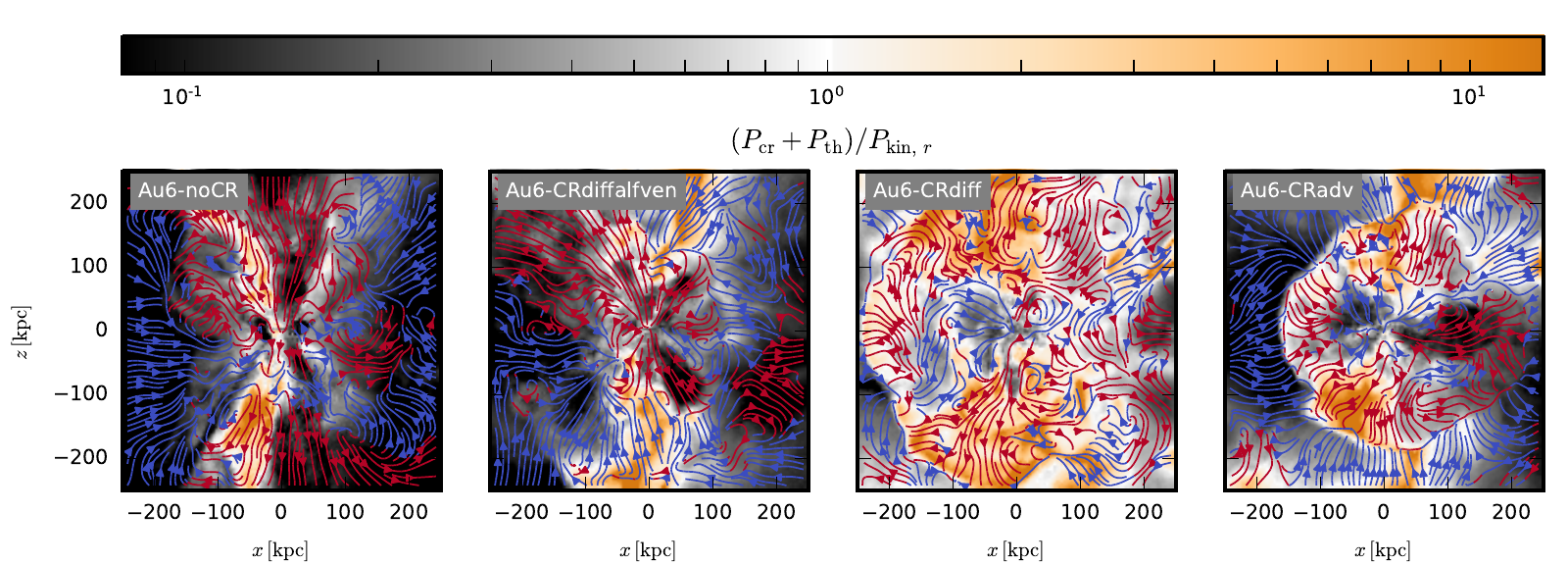}
\end{center}
\vspace{-.35cm}
\caption{CGM gas flow patterns of the halo Au6 at redshift $z=0.3$ where the stellar disc is shown edge-on. Streamlines indicate the direction of gas flow and arrow colors indicate the sign of the radial velocity of the flow. Red colors signal outflowing gas and blue colors in-flowing gas, respectively. The thickness of the gas slice is $20$ kpc in the $y$-direction and the background color-coding shows the pressure ratio $(P_\CR + P_\rmn{th})/P_{\rmn{kin},\,r}$, i.e., the sum of CR and thermal pressure over the radial kinetic flux term in the Euler equation, $P_{\rmn{kin},\,r}=\rho \varv_r^2$.}
\label{fig:flow}
\end{figure*}

\begin{figure*}
\begin{center}
\includegraphics[width=.49\textwidth]{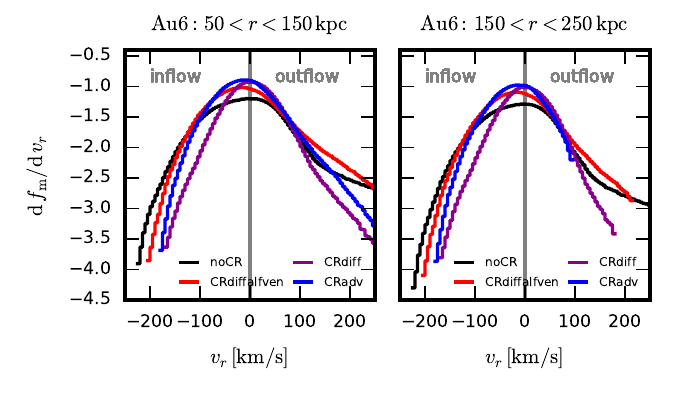}
\includegraphics[width=.49\textwidth]{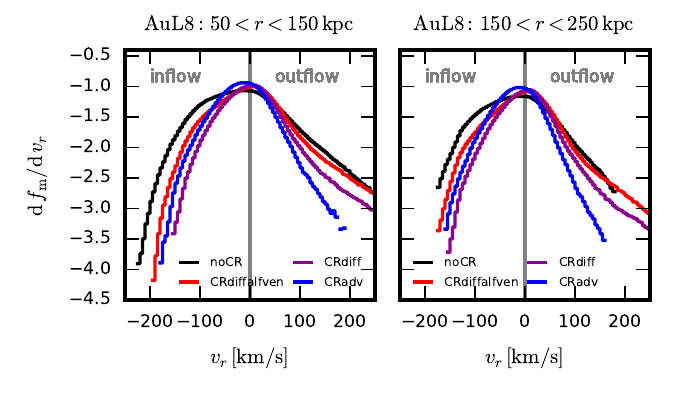}
\end{center}
\vspace{-.7cm}
\caption{Time averaged mass weighted distribution of the gas radial velocity in two different $100$ kpc wide spherical shells for the Au6 halo in the two left most panels and the Au8 halo in the two right most panels. The CR runs show a suppressed tail towards large inflow velocities in comparison to the noCR runs. Time averaging is done for 10 snapshots in the redshift range $0.2<z<0.3$ corresponding to a time span of $t\sim1.3$ Gyr.}
\label{fig:flow2}
\end{figure*}

\subsection{Gas accretion onto the disc}
\label{sec:accr}

We have seen that the inclusion of CRs lead to more compact stellar and gaseous discs. In this section, we investigate the evolution of the angular momentum of gas that ends up in stars in the central galaxy at present-day. In practice, we make use of Lagrangian ``tracer particles'' \citep{Genel2013,Grand2019} to follow the motion of resolution elements over time. At the beginning of each simulation, each gas cell in the high-resolution region is assigned a tracer particle with a unique ID. A tracer particle in any given cell moves to a neighbouring cell with a probability proportional to the outward mass flux across a cell face. Usually, a tracer particle has the highest probability  to remain in the same cell, because the moving-mesh nature of AREPO means that cells follow the bulk gas flow as closely as possible.  

Following the median angular momentum of gas which is in stars at redshift $z=0$ back in time (Fig.\ \ref{fig:ang_mom}) we find that CRs suppress the acquisition of angular momentum after the time of disc formation ($t_{\rm lookback}\sim5-8$ Gyr). The suppression is strongest for the CRadv and CRdiff runs while the CRdiffalfven run more closely follows the fiducial runs. At present-day, the CR simulations have acquired a factor of $\sim2-5$ times less specific angular momentum which manifests itself in more compact discs. For AuL8, the differences between the noCR run and the CR runs are smaller and the CRdiffalfven run matches the angular momentum of the noCR run. At a lookback time of $6$ Gyr this galaxy undergoes a merger which masks most of the differences in angular momentum distribution between the different physics runs. The evolution of the median angular momentum is indeed representative of the evolution of the entire distribution of gas angular momentum, as can be verified in Fig.~\ref{fig:ang_mom2}.

The angular momentum acquisition of the galaxy is most efficient if the accreted gas from large scales is undisturbed and flows to the central gas disc. When a large gas disc is first forming in the fiducial AURIGA model the wind feedback model develops outflows perpendicular to the stellar/gas disc (e.g. left panel in Fig.\ \ref{fig:flow}), which is an emergent phenomenon that is the result of the outflow taking the path of least resistance away from the galaxy \citep{Grand2019}. As we have discussed earlier (Fig.\ \ref{fig:gas}), in the CR runs the additional CR pressure support inflates the gas disc and thus enhances the gas density above and below the disc. The gas morphology in these runs is thus less discy and the particular implementation of the wind model results in more spherically symmetric gas flows that are less coherent in the perpendicular direction of the disc as we exemplify in Fig.\ \ref{fig:flow}. This figure shows the gas flow pattern in form of stream lines at redshift $z=0.3$ (corresponding to a lookback time of $\sim3.5$ Gyr).

In order to compare the wind properties to the dominant pressure forces and assess whether CRs change the hydrodynamic halo properties, we overlay the streamlines of Fig.\ \ref{fig:flow} on a colour map that shows the pressure ratio $(P_\CR + P_\rmn{th})/P_{\rmn{kin},\,r}$, where $P_{\rmn{kin},\,r}=\rho \varv_r^2$ is the radial kinetic flux term in the Euler equation. This can be seen by looking at at the momentum equation of an ideal fluid in the presence of CRs, which reads as follows:
\begin{equation}
  \frac{\partial \rho \bvel}{\partial t}
  + \bnabla\bcdot \left(\rho \bvel \bvel^\rmn{T} + P\mat{1} - \bB\bB^\rmn{T}\right)
  = -\rho\bnabla\Phi,
  \label{eq:euler}
\end{equation}
where $P=P_\CR + P_{\rmn{th}} + P_\rmn{mag}$ and $\mat{1}$ is the unit rank-two tensor. Converting this equation to spherical coordinates and neglecting the pressure and tension term of subdominant magnetic fields, the radial momentum flux density is given by $\rho \varv_r^2 + P_\CR + P_{\rmn{th}}$ \citep{Mihalas1984}. Figure \ref{fig:flow} shows that the divergence in the stream lines ($\bnabla \bcdot\bvel < 0$) corresponds to a shock where kinetic energy is converted into thermal energy.

The noCR simulation in Fig.\ \ref{fig:flow} shows coherent outflows along the direction of the spin axis of the disc, which enable flow channels to open up in the perpendicular direction along which low-metallicity gas can be coherently accreted to the central disc (Pakmor et al. in prep.). Whereas the gas flow in the CRdiffalfven run more closely resembles the flow pattern of the fiducial AURIGA model, the direction of the outflows is less ordered and not always perpendicular to the disc. The CRdiff and CRadv runs on the other hand show more spherical, slower outflows which shock the inflowing gas at a distance of $r\sim100-200$~kpc, shutting off the coherent gas inflows to the central galaxy while at the same time preventing coherent outflows to large distances. Therefore, these models exhibit a more quiescent, hydrostatic atmosphere in the halo in comparison to the former two models as can be see from the larger ratios of $(P_\CR + P_{\rmn{th}})/P_{\rm{kin,}\, r}$ (yellow colors in Fig. \ref{fig:flow}). We verified that this result remains qualitatively similar over the entire redshift range $0\leq z\lesssim1$ and we quantified the hydrodynamic effect of CRs on the gas flow with the distribution of radial inflow and outflow velocities in Fig.\ \ref{fig:flow2}.

The simulations with CRs show narrower radial velocity distributions indicating reduced inflow and outflow velocities. This is the result of the process described above: the more elliptical and vertically expanded ISM in the CR runs precludes a geometrically preferred path of least resistance and slows down the outflows in all directions. Hence there are no coherent outflows forming along the spin axis in the CR simulations. Because there are low-velocity outflows present in nearly all directions in these CR runs, this shocks the accreting gas and precludes the formation of most inflow channels that deliver gas from larger distances to the star forming disc. The suppression of the inflow velocities is strongest for the CRdiff run because here CRs impact a larger region compared to the other runs. Interestingly, in the CRdiffalfven simulation we observe reduced infall velocities but similar or even larger outflow velocities in comparison to the fiducial AURIGA model due to the additional CR pressure-driven winds.

\begin{figure}
\vspace*{-.5cm}
\begin{center}
\includegraphics[width=1.2\columnwidth]{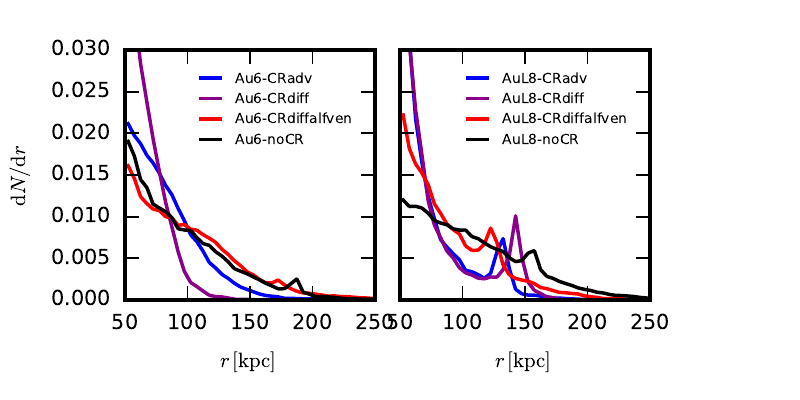}
\end{center}
\vspace{-.75cm}
\caption{Radial distribution at a lookback time of $5$ Gyr of the gas tracers which at present-day are contained in stars. This figure traces the origin of the gas that evolves into stars at the present day. For AuL8 in the right panel, the peaks around $R\sim140$ kpc signal an ongoing merger.}
\label{fig:dist}
\end{figure}

The immediate manifestation of this process is the suppression of the accretion of gas from larger distances in the CR runs as displayed in Fig.\ \ref{fig:dist}. In this figure we show the distribution of radial distances at a lookback time of $5$ Gyr of the gas which is at present-day converted into stars of the stellar disc. In the fiducial AURIGA runs gas is accreted from farther away in comparison to the CR counterpart simulations. This is the result of the modified gas accretion pattern on large scales mediated by the effects of the CRs on the structure of the gaseous disc on smaller scales. Note, this does not necessarily mean that SFRs in the CRdiff and CRadv run are suppressed as the outflow velocities are similarly reduced in these runs leaving enough gas to fuel star formation.

In summary, in the fiducial AURIGA and the CRdiffalfven simulations gas accretes relatively unimpeded from large distances whereas in the CRadv and CRdiff simulations the more spherically symmetric outflows and the CR pressurised gaseous haloes are able to held up the gas.

\section{Circum-galactic medium} \label{sec:CGMprops}

\begin{figure*}
\begin{center}
\raggedleft
\includegraphics[width=\textwidth]{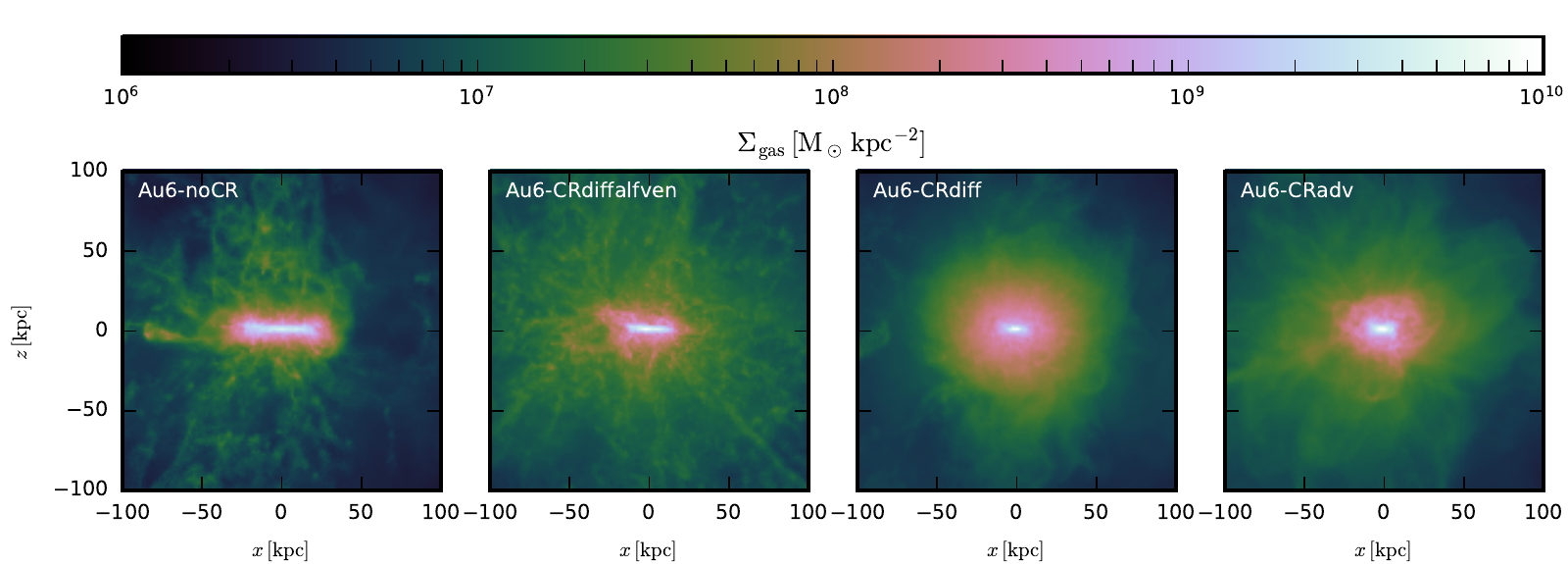}
\includegraphics[width=\textwidth]{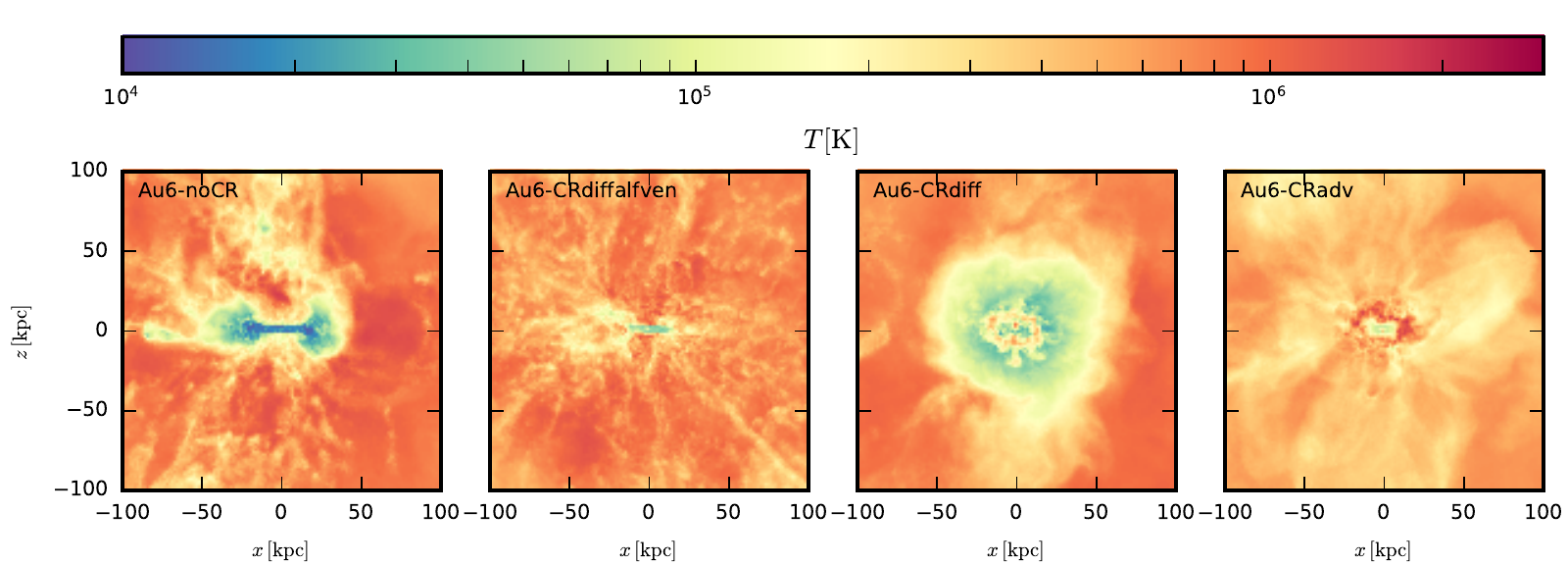}
\includegraphics[width=.875\textwidth]{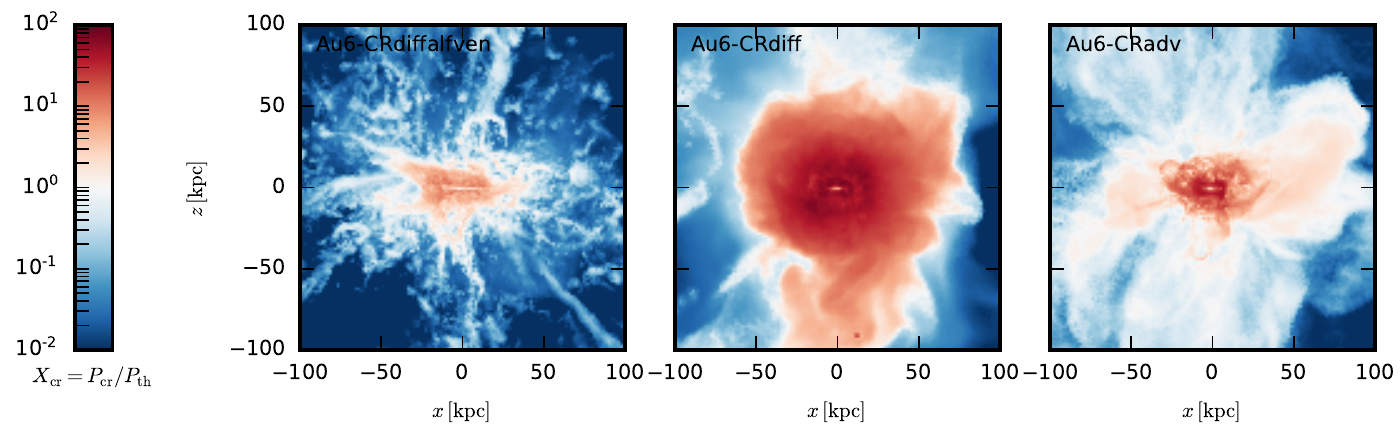}
\includegraphics[width=.875\textwidth]{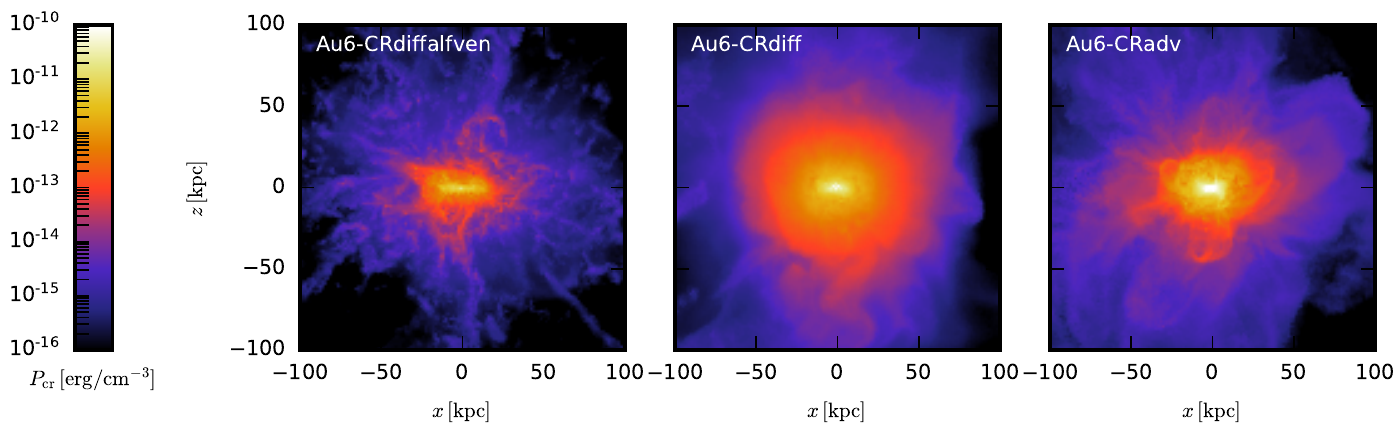}
\end{center}
\vspace{-.35cm}
\caption{Maps of CGM properties for the four different physics runs of halo Au6 as indicated in each panel. From top to bottom we show the gas surface density, the gas temperature, the CR-to-thermal pressure ratio and the CR pressure. The orientation of each panel is chosen to view the central disc edge-on and the projection depth of each panel is equal to its width, $200$ kpc. Note the smooth gas distribution in the CR runs owing to the additional pressure of the CRs, which however differs considerably for our different variants of CR transport.}
\label{fig:CGMgas}
\end{figure*}

\begin{figure*}
\vspace*{-.3cm}
\begin{center}
\includegraphics[width=\textwidth]{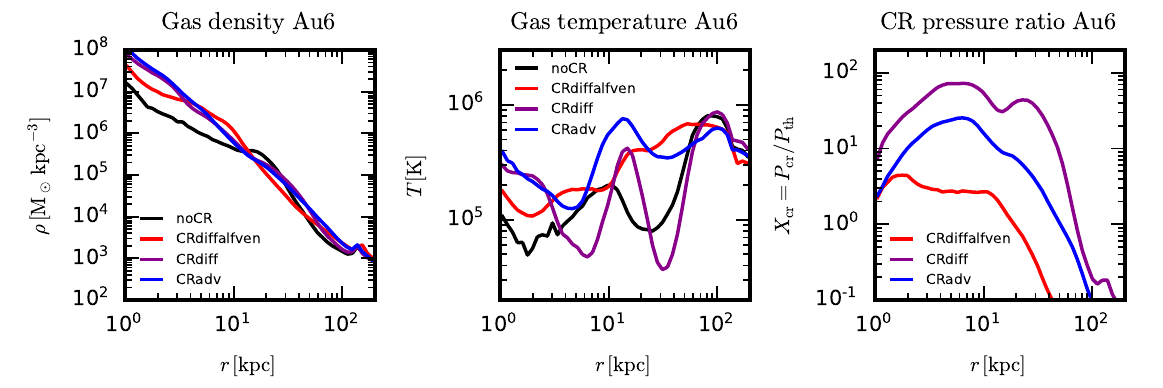}\vspace{-.15cm}
\includegraphics[width=\textwidth]{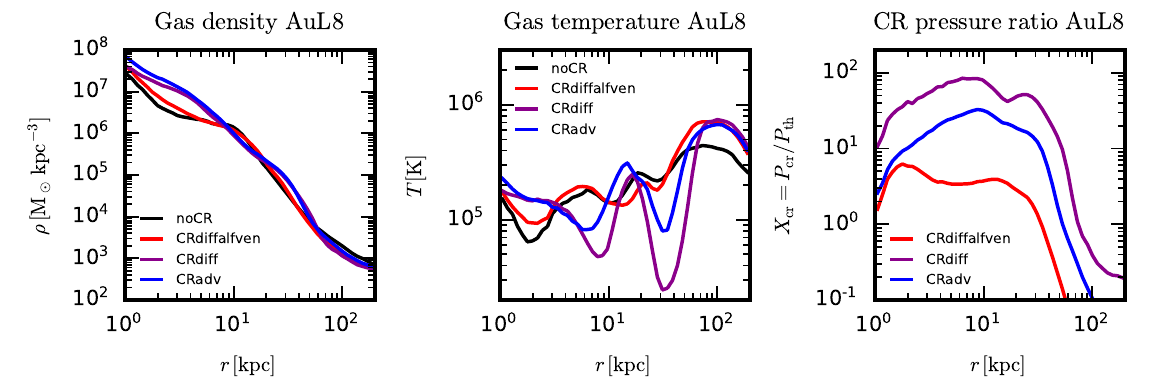}
\end{center}
\vspace{-.35cm}
\caption{Comparison of the radial profiles of the gas density (left), gas temperature (middle panel) and magnetic field strength (right) in the four different physics variants of Au6 (upper panel) and AuL8 (lower panels). The profiles represent time averages over the last 10 simulation outputs ($z\sim0.1$, $t\sim1$ Gyr).}
\label{fig:CGMprof}
\end{figure*}

We now turn to analyse the effects of CRs on the CGM properties in the different physics variants. In particular, we focus on how CRs shape the gas density distribution and impact the gas temperature profile of CGM gas by providing additional pressure support which manifests itself in high ratios of CR pressure to thermal pressure. To this extent we show in Fig.\ \ref{fig:CGMgas} from top to bottom maps of the gas surface density, the gas temperature, the ratio of CR pressure to thermal pressure and the value of CR pressure for galaxy Au6. The orientation is chosen such that the gas disc is seen edge-on and the projection depth is the same as the horizontal/vertical extend ($200$ kpc). For a more quantitative comparison we accompany the maps by profiles of the same quantities for both galaxies in Fig.\ \ref{fig:CGMprof} as indicated by the panels' titles. For the CGM properties we have chosen logarithmically spaced spherical shells and averaged profiles over the last ten simulation outputs ($\sim1$ Gyr). Color-coding of the different physics variants is the same as in previous figures.

Looking at the first row of Fig.\ \ref{fig:CGMgas} and comparing the four different physics variants, we find that the CGM gas surface density in the CR runs outside the disc region ($R>50$ kpc) is slightly higher in comparison to the fiducial AURIGA model. Most strikingly, the CGM gas density is significantly more spherical within $50$ kpc in the CRdiff and CRadv runs compared to the fiducial AURIGA  and the CRdiffalfven runs. Additionally, the CGM gas density is smoother in the CR runs in comparison to the noCR run. In the next subsections we address these morphological differences and highlight how CRs cause these changes of the CGM structure by investigating each model separately.

\subsection{AURIGA -- no cosmic rays}
\begin{figure*}
\vspace*{-.4cm}
\hspace{-.4cm}
\includegraphics[width=.515\textwidth]{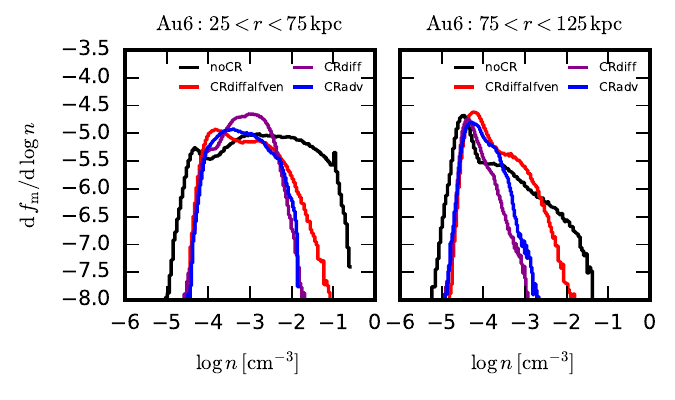}
\hspace{-.4cm}
\includegraphics[width=.515\textwidth]{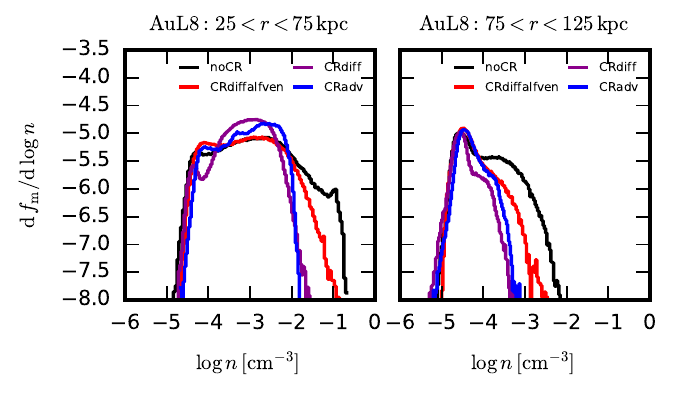}
\vspace{-.35cm}
\caption{Distribution of gas density in two different $50$ kpc wide spherical shells for the Au6 halo in the two left most panels and the Au8 halo in the two right most ones. The CRadv and CRdiff runs exhibit narrower density distributions in comparison to the other two models indicating a smoother CGM.}
\label{fig:CGMhist}
\end{figure*}

The baseline model for our comparison is the fiducial AURIGA model which has a highly structured CGM with cool, high density patches coexisting next to hot low density regions (left panels in Fig.\ \ref{fig:CGMgas}, see also \citealt{Vandevoort2019}). The clumpy CGM morphology is further reflected in a broad gas density distribution with tails to large gas densities as shown in Fig.\ \ref{fig:CGMhist}. Here we show the density distribution of all four runs in two different concentric shells of width 50 kpc as indicated by the panel titles. The left two panels show galaxy Au6, the right two AuL8.

\subsection{Cosmic ray advection}
The most simple approximation for CR transport physics is the advection of CRs, which neglects all active CR transport processes. Therefore, all CRs in the CGM have been transported there by outflows. In the CRadv run the CGM is significantly more spherical within $50$ kpc, smoother and of slightly lower temperature (right panel of Fig.\ \ref{fig:CGMgas}) when compared to the fiducial AURIGA run owing to the additional CR pressure. Especially far away from the disc at $R\sim50$ kpc the density is slightly enhanced in comparison to the fiducial run. The inclusion of CRs leads to cooler gas temperatures even at distances of $R\sim100$ kpc where CR pressure is approximately in equilibrium with the thermal pressure. The additional CR pressure smoothes out almost all small-scale high density peaks in the CGM gas which quantitatively leads to a narrower gas density distribution in Fig.\ \ref{fig:CGMhist}.

\subsection{Cosmic ray diffusion}
Allowing for CR anisotropic diffusion alters the properties of the CGM dramatically but with similarities to the CRadv run. We find that the CGM gas density is even more spherical within $50$ kpc compared to the CRadv run, highly CR pressure dominated and of much cooler temperatures. In this run, CRs are allowed to diffuse and thus are able to affect the CGM at larger distances from the disc, thus the CR pressure dominated halo is larger in size compared to the CRadv run. Furthermore, the CR pressure contribution is higher compared to the CRadv run (see also discussion in Sections \ref{subsec:gasdisc} and \ref{subsec:dis}) leading to an even smoother CGM (see narrow gas distribution in Fig.\ \ref{fig:CGMhist} for this run). The additional CR pressure which in this run dominates the CGM out to radii of $R\sim(50-100)$ kpc supports the gas against gravitational collapse in the absence of thermal pressure support and thus explains the low CGM temperatures which coincide with the regions where CR pressure dominates. Figure~\ref{fig:CGMprof} shows that in a region of $R<50$ kpc the CR pressure is a factor of $\sim10$ larger than the thermal pressure.

\begin{figure*}
\begin{center}
\includegraphics[width=\textwidth]{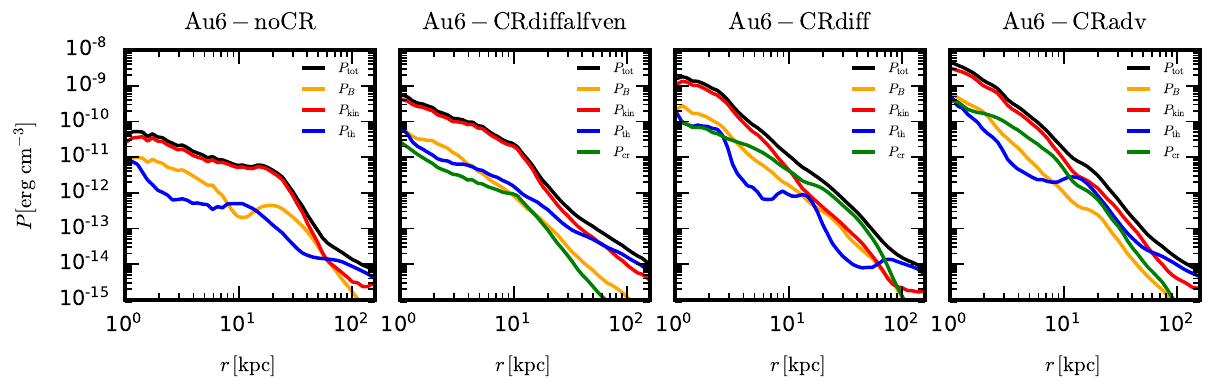}
\includegraphics[width=\textwidth]{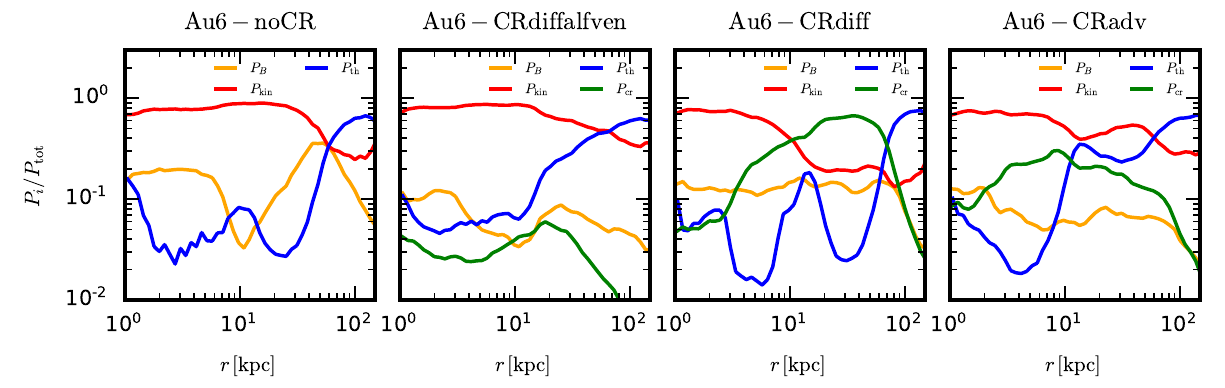}
\end{center}
\vspace{-.35cm}
\caption{Radial profiles of different pressure components for the four different physics variants of halo Au6. The upper panels compare absolute pressure profiles of the magnetic ($P_{B}=\bB^2/(8\pi)$, orange line), ``kinetic'' ($P_{\rm kin}=\rho \bvel^2/2$, red), thermal ($P_{\rm th}=(\gamma-1)\eps_{\rm th}$, blue) and CR ($P_{\rm cr}=(\gamma_{\rm cr}-1)\eps_{\rm cr}$, green) components as well as the total pressure (black). Lower panels compare the relative pressure contributions to the total pressure.}
\label{fig:pressure}
\end{figure*}

\begin{figure*}
\vspace{-.25cm}
\begin{center}
\includegraphics[width=\textwidth]{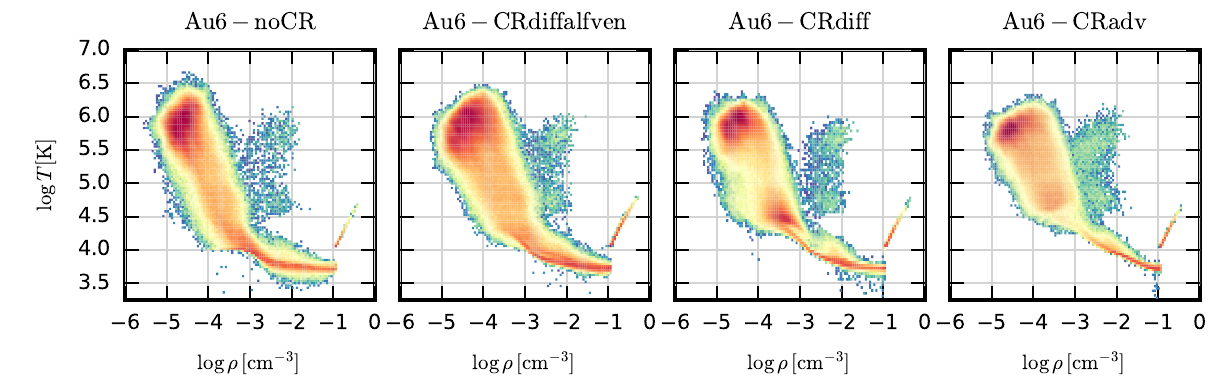}
\end{center}
\vspace{-.35cm}
\caption{Present-day temperature-density phase diagrams of CGM gas ($50<R<200$ kpc) for the four different physics runs of galaxy Au6. Color-coding shows the logarithm of the mass weighted probability density where red colors indicate high and blue colors low probability.}
\label{fig:phase}
\end{figure*}

\subsection{Cosmic ray diffusion with Alfv\'en wave cooling}
The CRdiffalfven model in turn reveals a CGM morphology similar to the fiducial AURIGA model (although with noticeable differences) and shows clear differences in comparison to the CRdiff and CRadv run. In comparison to the AURIGArun (CR runs) the CGM features smoother (more structured) density peaks and the central CGM appears discy. Again, the additional CR pressure explains the smoother gas density compared to the AURIGA model while the Alfv\'en cooling explains the weaker damping compared to the CRdiff and CRadv runs. Therefore in Fig.\ \ref{fig:CGMhist}, the gas density distribution of the CRdiffalfven run lies between the AURIGA model and the other two CR models. 

In comparison to the CRdiff run the CGM is much hotter and shows even less cold regions compared to the fiducial run. In fact, the profiles (middle panels in Fig.\ \ref{fig:CGMprof}) show that Au6 has a hotter CGM in the CRdiffalfven run compared to the AURIGA run. AuL8 on the other hand shows a similar CGM temperature. The slightly enhanced CGM temperature in Au6 is presumably due to the additional Alfv\'en heating of the CRs and the fact that Au6 had a recent burst of SF injecting CRs into the CGM. This burst is not present in AuL8. From the lower panels in Fig.\ \ref{fig:CGMgas} we can further see that the CR pressure dominates the central regions (except for the disc) while for most of the CGM gas the CR pressure is in equilibrium with the thermal pressure (see also right panels in Fig.\ \ref{fig:CGMprof}). This is different to the other CR runs and a manifestation of the Alfv\'en wave cooling in which the CRs lose an e-folding of their initial energy as they diffuse a scale height into the CGM. 

\subsection{CGM pressure support}
In Fig.\ \ref{fig:pressure} we compare in detail the different pressure components (upper panels) and their contribution to the total pressure (lower panels) in the CGM. We compare magnetic pressure (orange), ``kinetic pressure'' (red), thermal pressure (blue) and CR pressure (green) to the total pressure (black) in spherical shells as we have explained for Fig.\ \ref{fig:CGMprof}. From left to right we show the fiducial AURIGA noCR, the CRdiffalfven, the CRdiff and the CRadv runs. 

In the region influenced by accretion onto the disc as well as the disc itself (within a few tens of kpc) the gas is rotationally supported (i.e., has a dominating kinetic pressure) in all runs while the other pressure components, i.e. thermal, magnetic and CR pressure, are roughly in equipartition contributing each about $\sim10$\% to the total pressure. In the outskirts at radii larger than $R>20$ kpc we find that all the runs become increasingly thermal pressure dominated except for the CRdiff run where CR pressure dominates and the thermal pressure becomes negligible. Only close to the virial radius the thermal pressure becomes important again which was already noted by the huge CR pressure dominated halo in Fig.\ \ref{fig:CGMgas}. This confirms our previous findings that the cool region in the CGM at these radii is entirely CR pressure dominated.

\subsection{Temperature-density relation}
Our findings for the structure and morphology of the CGM are summarized in Fig.\ \ref{fig:phase} showing the temperature-density distribution of the CGM gas in the radial range $50<R<200$ kpc. While at first glance differences between the fiducial run and the CR runs are small, one notices that the hot phase in the CR runs tends to inhabit regions of lower temperature. In more detail, we find that the CRdiffalfven runs show a larger spread in temperature at any density compared to the noCR run which shows the importance of CR Alfv\'en heating. This figure further shows that for the CRdiff run gas at $\rho\sim(10^{-4} - 10^{-3})$~cm$^{-3}$  piles up at a temperature of $T\sim10^{4.5}$ K. We interpret this as the CR pressure keeping this gas from falling onto the main galaxy. Finally, we find that in the CRadv run the CR pressure causes a different slope of the $\rho-T$ relation for the non-starforming gas at $T\sim10^{4}$ K due to the CR pressure support at the disc-halo interface.

Thus, CRs do not only affect the properties of the gas disc as we have seen in Fig.\ \ref{fig:gas} but also the gas morphology of the CGM even at large distances close to the virial radius of the halo. The analysis in this section reinforces the need to better understand the physics of CR transport, as we have shown here that different variants of approximating it have a strong impact on the stellar structure and especially the properties of the CGM.

\section{Far-infrared$-$gamma-ray relation}
\label{sec:obs}

\begin{figure}
\begin{center}
\includegraphics[width=\columnwidth]{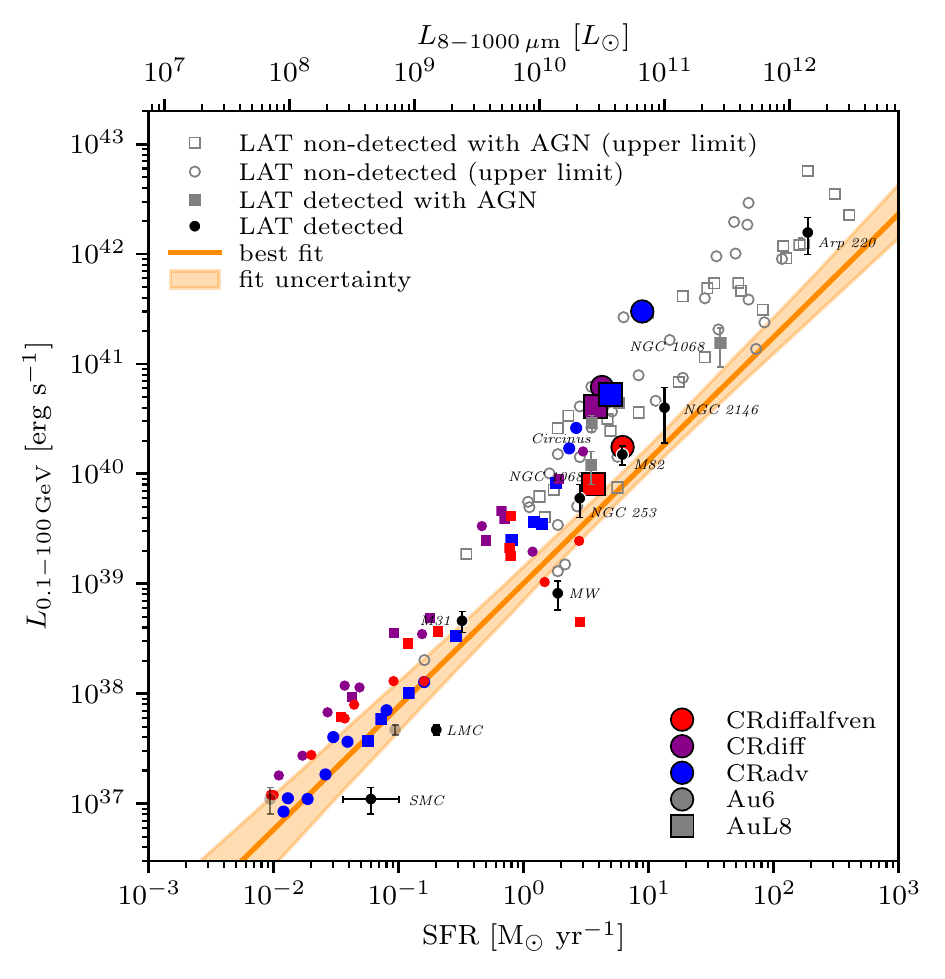}
\end{center}
\vspace{-.35cm}
\caption{Correlation of the gamma-ray luminosity ($L_{0.1-100\ \rm GeV}$) with the SFR and with the FIR luminosity ($L_{8-1000\ \rm mm}$) of star-forming galaxies. We compare our six simulated CR models (differently coloured symbols delineate Au6 and AuL8 haloes) and plot central galaxies (thick symbols) and dwarf galaxies (thin symbols). Upper limits on the observable gamma-ray emission by Fermi-LAT (open gray symbols; \citep{Rojas-Bravo2016}) are contrasted with gamma-ray detections from star-forming galaxies only (solid black) and with AGN emission (filled gray); data are taken from \citet{Ackermann2012}, except for NGC 2146 \citep{Tang2014} and Arp 220 \citep{Griffin2016,Peng2016}.
For the two lowest SFR galaxies, the SMC and LMC we use gray data points for the observed FIR luminosities (top axis) and solid black data points for the SFRs derived using H$_\alpha$ and UVBI photometry.
Note that only the CRdiffalfven runs fall on the best-fit observational FIR–gamma-ray correlation (orange).}
\label{fig:gamma}
\end{figure}

Finally, after establishing the differences and similarities between the three variants of CR transport in our simulations we connect our results to the most directly observable CR proton properties of galaxies, namely hadronic gamma-ray emission that arises from inelastic collisions of CRs with the ambient ISM. To this end we compare in Fig.\ \ref{fig:gamma} the gamma-ray luminosities in the Fermi band ($0.1-100$GeV) to the SFR for all the main disc galaxies (big black bordered symbols) in comparison to observational data as indicated in the caption. Additionally, we also show gamma-ray luminosities for the dwarf galaxies within the zoom region (small coloured dots). Note that these dwarfs are not satellite galaxies as they are not part of the main halo but proxies of field dwarfs in the Local Volume. Note that our homogeneous observational sample in the Fermi band spans an energy range of $0.1-100$~GeV leading to somewhat higher gamma-ray luminosities in comparison to previous observational data collections \citep[cf.][who use the smaller energy band between $1-100$~GeV]{Lacki2011}.

The total far-infrared (FIR) luminosity ($8-1000~\mu$m) is a well-established tracer of the SFR of spiral galaxies \citep{Kennicutt1998} with a conversion rate \citep{Kennicutt1998ARA+A} 
\begin{equation}
  \label{eq:FIR-SFR}
  \frac{\rmn{SFR}}{\rmn{M}_\odot~\rmn{yr}^{-1}}=\epsilon\,1.7\times10^{-10}\,\frac{L_{8-1000\,\mu\rmn{m}}}{L_\odot}.
\end{equation}
This SFR-FIR conversion assumes that thermal dust emission is a calorimetric measure of the radiation of young stars, and the factor $\epsilon=0.79$ derives from the \citet{Chabrier2003} IMF \citep{Crain2010}. While this conversion is reliable at $L_{8-1000\,\mu\rmn{m}}>10^9~L_\odot$, it becomes progressively worse at smaller FIR luminosities due to the lower metallicity and dust content, which implies a low optical depth to IR photons and invalidates the calorimetric assumption \citep{Bell2003}. We have verified that down to SFRs comparable to those of M31 ($\sim0.3-0.4$~\Msun~\rm{yr}$^{-1}$) our conversion still holds. The SFR of M31 derived here is in good agreement with SFRs derived using a combination of H$\alpha$ and 24~$\umu$m emission, a combination of far-ultraviolet, 24~$\umu$m, and the total infrared emission which yield $\sim(0.35-0.4)\pm0.04$~\Msun~\rm{yr}$^{-1}$ \citep[see][]{Rahmani2016}.

We show the observed FIR luminosity of the LMC and SMC with gray data points in Fig.\ \ref{fig:gamma}, while SFR estimates for the LMC and SMC are shown with solid black data points. Those are derived by combining H$_\alpha$ and FIR emission \citep[assuming a Chabrier~IMF,][]{Wilke2004} or UVBI photometry \citep{Harris2009} and range for the SMC between 0.036 and 0.1 \Msun yr$^{-1}$, while the LMC forms $0.2~\Msun~\rmn{yr}^{-1}$ of stars \citep{Harris2009}. We refer the reader to \citet{Pfrommer2017b} for more details on the FIR-to-SFR conversion and to \citet{Pfrommer2004} for the computation of the gamma-ray emission resulting from hadronic proton interactions with the ambient ISM.

We find that at the MW mass scale or vice-versa at a FIR luminosity of $\sim5\times10^{10}$ L$_{\odot}$ the three CR models predict very different gamma-ray luminosities. Whereas the CRdiffalfven runs are in good agreement with the observational data, especially for NGC253 and M82, the other two models lie above the best fit observational relation of \citet{Rojas-Bravo2016} and are barely consistent with upper limits from Fermi-LAT (gray open symbols). This is interesting as the same physical models for CR transport (CRadv and CRdiff) yield hadronic gamma-ray luminosities in agreement with the best fit observational relation for the collapsing halo initial conditions \citep{Pfrommer2017b}. Unlike our cosmological simulations, these idealized setups produce more extended gas and stellar discs. As a result the inner gravitational potential due to the kinematically cold distribution of central stars is shallower in these idealized models in comparison to our cosmological simulations in the CRdiff and CRadv models. Such a massive central stellar distribution in our cosmological simulations injects too many CRs and compresses the gas to a level that the CRs overproduce the hadronic gamma-ray emission. Clearly, different types of feedback (radiation, supernovae) are necessary to act in tandem with CRs to prevent the formation of these dense cores at high redshift in our cosmological simulations.

At lower SFRs and accordingly lower FIR luminosities, the differences between the three models become smaller and both the CRdiffalfven and the CRadv model agree with the observed relation of \citet{Rojas-Bravo2016} while the CRdiff model produces a somewhat elevated level of gamma-ray luminosity. We attribute this behaviour to the lower injection rate of CRs at these SFRs and higher escape fractions of CRs from these low mass galaxies \citep[][]{Zhang2019}. Our CRdiffalfven model is able to explain the distribution of luminosities of the MW and M31. In particular at SFRs comparable to M82 and NGC~253 only this model matches observed gamma-ray luminosities. 

However, the simulations of all CR models in Fig.~\ref{fig:gamma}, including the CRdiffalfven model, do not agree with the observed SFRs and gamma-ray luminosities of the LMC and SMC. This can either be interpreted as an overprediction of the simulated gamma-ray luminosity (by a factor of about $2-2.6$ for the LMC and about $6-8$ for the SMC depending on the specific simulated counterpart), an underprediction of the simulated SFR (by a factor of $2-10$ for the LMC and $2-6$ for SMC), or a combination of both. There are several reasons that could be responsible for this. Among those is the omission of CR streaming, which will be studied in a forth-coming paper. In fact, CR streaming in combination with diffusion might lead to an increased effective CR transport speed which might decrease the gamma-ray luminosity in dwarf galaxies \citep[e.g.][]{Salem2016,Jacob2018,Chan2019}. Moreover, we would like to emphasise that the LMC and SMC are satellite galaxies of the MW, which are most likely at their first infall and in an interacting state \citep{Harris2009}. This might lead to higher SFRs in comparison to the isolated, non-satellite dwarf galaxies studies in Fig.\ \ref{fig:gamma}. While the CR production scales with the SFR and accordingly the gamma-ray luminosity, it is a matter of time-scales involved whether the gamma-ray luminosity closely follows the measured SFR \citep[which is intrinsically uncertain by a factor of two for the LMC, see e.g.,][]{Harris2009}. In fact the LMC SFR shows a strong burst in the last few $10^7$ yr \citep[cf.\ Fig. 11 of][]{Harris2009}. For such a recent burst we do not expect that the gamma-ray production had time to fully react as the lifetime of galactic CRs is of the order of $\gtrsim3\times10^7$ yr \citep[e.g.,][]{Simpson1988,Lipari2014}.

Finally, our simulated points agree with the mean power-law relation. It is unclear whether the true observational relation continues along the power law relation with the LMC and SMC representing outliers or whether the lower gamma-ray luminosities of the LMC and SMC indeed signal a cutoff or change in slope of this relation.

We note that at low SFRs the CRadv model produces the lowest gamma-ray luminosities while at high SFRs this model shows the opposite effect. This outcome can be explained by a different behaviour of the adiabatic processes in halos of different masses. Figure~4 of \citet{Pfrommer2017} shows that adiabatic losses dominate in smaller halos corresponding to lower SFRs while adiabatic gains prevail in larger halos. Especially, at low SFRs the adiabatic losses in the CRadv model are larger compared to those in the CRdiff model which explains the low gamma-ray luminosities of this model at low SFRs.

Note that our CR models reproduce the observed relation for Galactic values of the diffusion coefficient ($\kappa_\parallel=10^{28}\,\rmn{cm^2~s}^{-1}$) or even without CR diffusion (for our CRadv model). This is in stark contrast to the FIRE-2 simulations that require diffusion coefficients of $\kappa_\parallel>3\times10^{29}\,\rmn{cm^2~s}^{-1}$ to be consistent with the gamma-ray observations \citep{Chan2019,Hopkins2019}. These differences can be traced back to the ISM model of AURIGA, which supports a CR transport with a single diffusion coefficient while the highly structured multi-phase ISM of FIRE-2 would require to fully model CR streaming with various wave damping mechanisms that dominate in the cold (ion-neutral damping) and warm/hot phases (non-linear Landau damping), respectively.

In conclusion, observable scaling relations such as the gamma-ray-FIR relation offer promising tools for distinguishing physically valid models and might help to constrain the values of free sub-grid model parameters \citep[e.g.][]{Buck2019}. However, a detailed model comparison to observations like, e.g., the CGM properties derived from the COS-HALOS survey \citep{Tumlinson2013} needs to include careful post-processing of the simulations which is outside the scope of this study and thus left for future research.

\section{Discussion} \label{sec:dis}

\begin{figure*}
\begin{center}
\vspace*{-.55cm}
\includegraphics[width=1.06\textwidth]{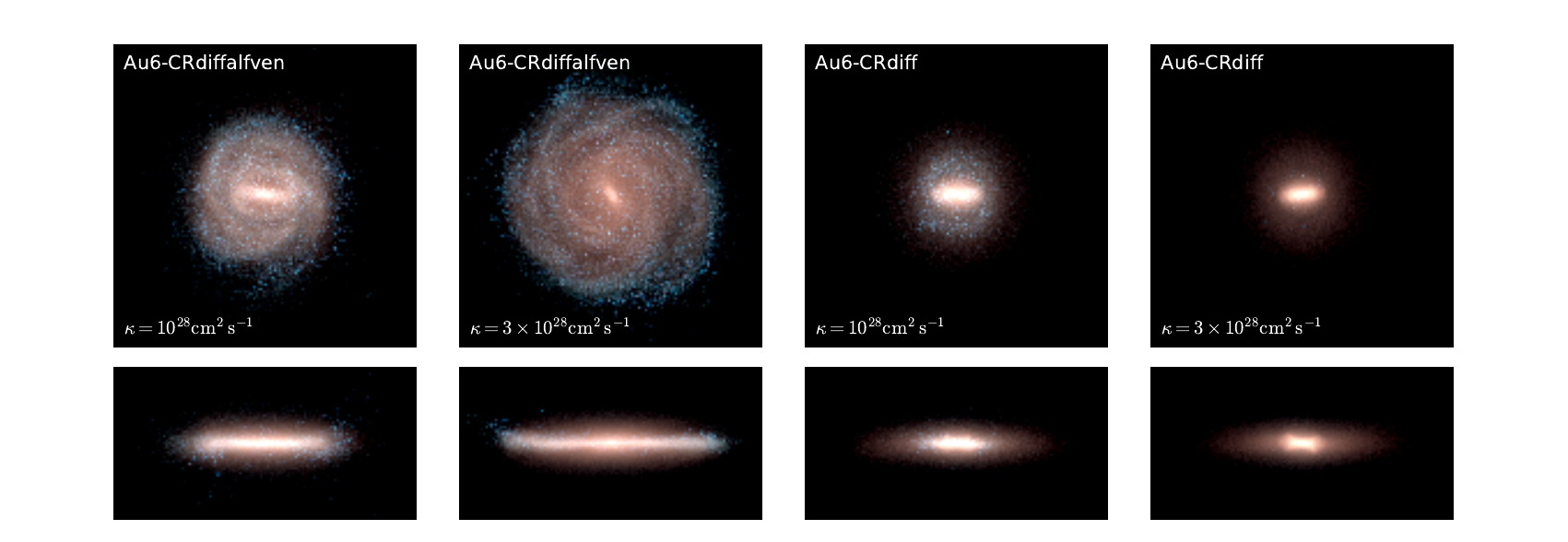}
\includegraphics[width=\textwidth]{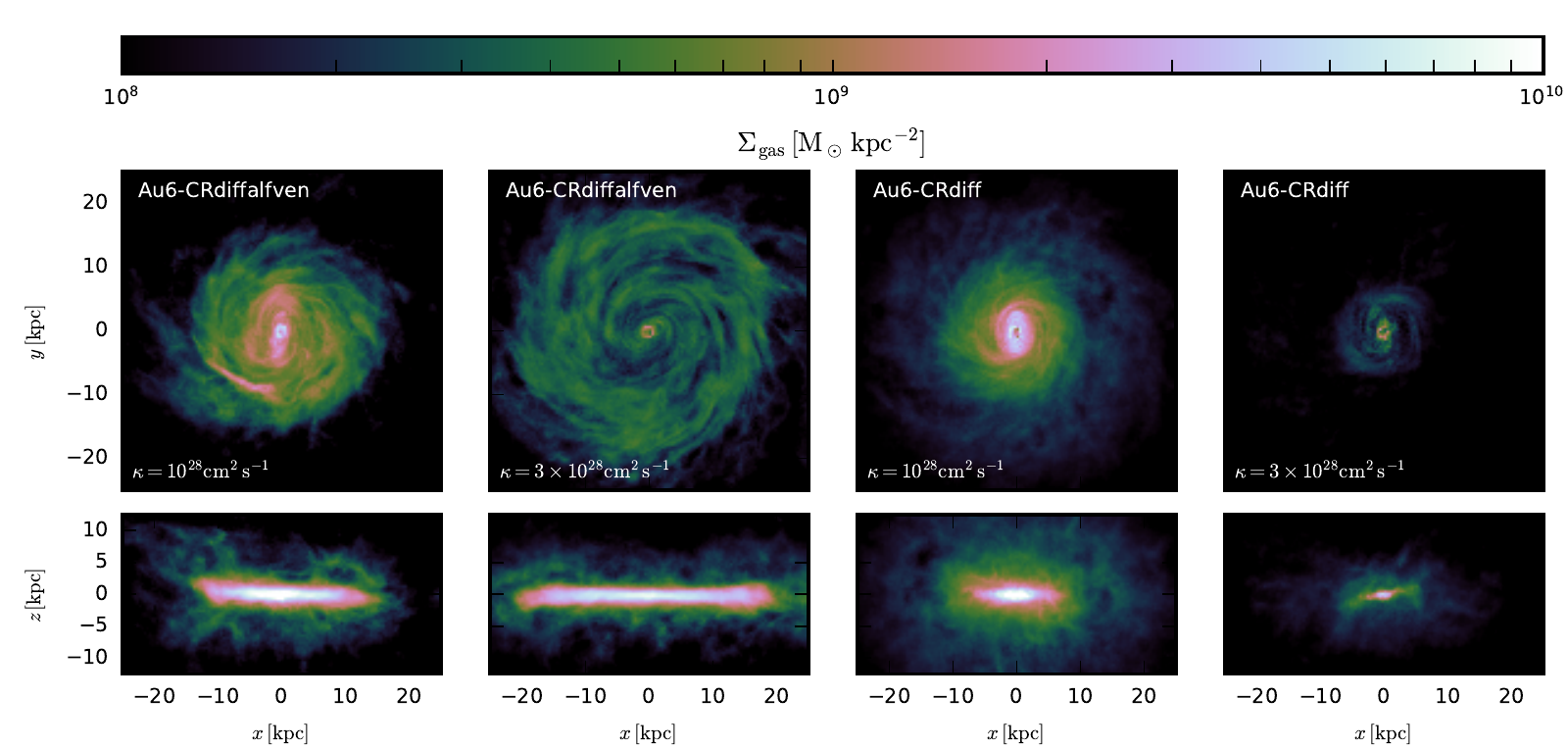}
\includegraphics[width=\textwidth]{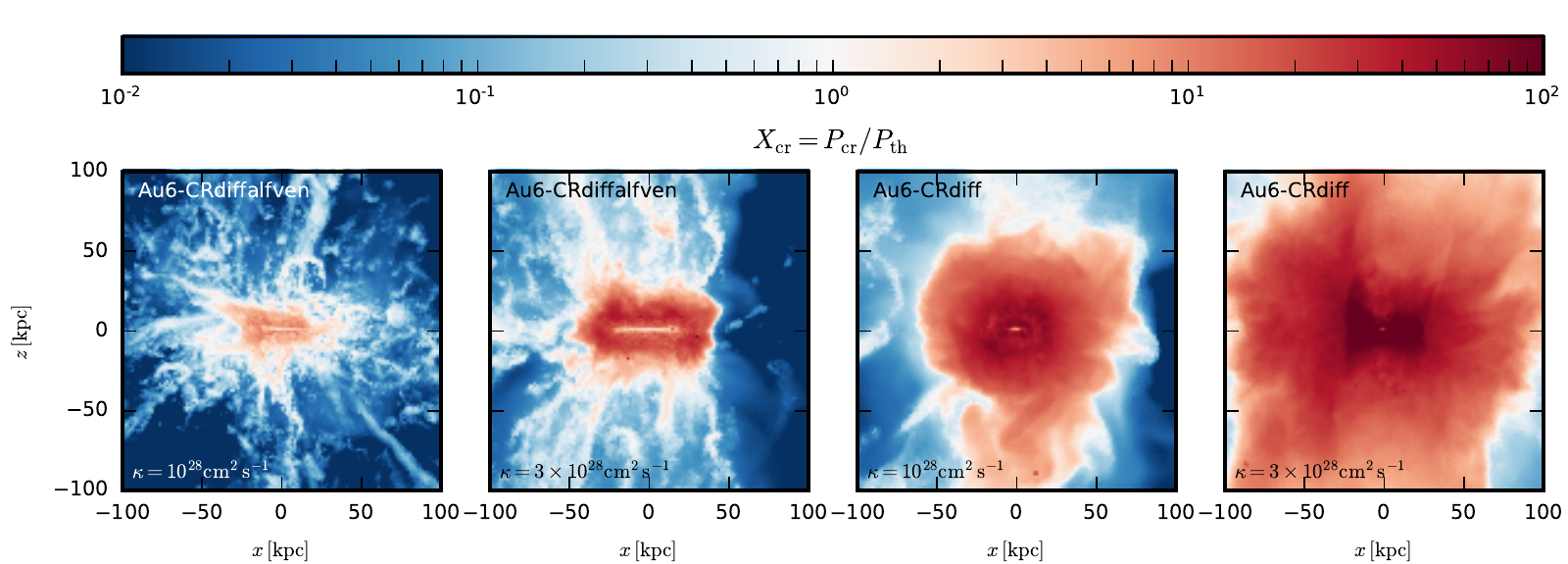}
\end{center}
\vspace{-.35cm}
\caption{Face-on and edge-on projection of the stellar light at $z=0$ (upper panels), the gas surface density (middle panels), and the CR-to-thermal pressure ratio, $X_{\rmn{cr}}$ (bottom panels), for the CRdiffalfven and the CRdiff model for two different diffusion coefficients as indicated in the panels. See caption of Fig. \ref{fig:rgb} and \ref{fig:gas} for further details.}
\label{fig:high_kappa}
\end{figure*}

\subsection{Comparison to observations}
\label{sec:obs_comp}

For galactic CRs there exist three different observables which can be used for scrutinising our model predictions: the local energy density of CRs, the gamma-ray luminosity and the CR grammage, $X_s=\int_\rmn{source}^\rmn{observer}\rho_{\rm nuclei}{\rm{d}} l_{\rm CR}$. In this paper, we have calculated the former two observables because those can be robustly derived from our models whereas the latter quantity is a gyro-radii averaged integration of the target density along the path of individual CRs and strongly depends on a number of properties. Among those are the exact initial CR energies that are mapped onto the observed CR spectra, the CR source distribution, the variability of CR injection (both in time and space), the exact topology of the magnetic field (at large and small scales, which are necessarily unresolved in current cosmological simulations and would have to be modelled sub-grid) as well as the exact small scale density distribution in the ISM along the CR path. Therefore, deriving a meaningful value for the CR grammage in our simulations is beyond the scope of this paper and we leave such a detailed comparison for future studies.

Comparing our model observables for the local energy density of CRs (right panels of Fig.\ \ref{fig:prof}) we find that the CRdiffalfven model results in the lowest CR energy densities.
At the solar circle ($r_{xy}\sim8$ kpc) we find a value of $\epsilon_{\rm cr}\sim2$ eV~cm$^{-3}$ for AuL8 and $\epsilon_{\rm cr}\sim4-5$ eV~cm$^{-3}$ for Au6. A fair comparison to MW measurements needs to account for the fact that both model galaxies show slightly higher SFRs
of $\sim2$ for AuL8 and $\sim3$ for Au6 (see e.g., Fig.\ \ref{fig:gamma}) 
in comparison to MW rates of $\sim1.65-1.9$ \Msun yr$^{-1}$ \citep[e.g.,][]{Chomiuk2011,Licquia2015}. Because the SFR is directly proportional to the CR energy input in our models, we expect approximately 1.3 to two times lower CR energy densities in those models at SFRs comparable to the MW. This leaves us with CR energy densities of $\sim1.6$ eV~cm$^{-3}$ for AuL8 and $\sim2-2.7$ eV~cm$^{-3}$ for Au6. Given that inferred observational results are uncertain by a factor of $\sim2$ \citep[see e.g. discussion in][]{Cummings2016} we conclude that the rescaled CR energy densities of those models are in reasonable agreement with estimates for the MW of about $\epsilon_{\rm cr}\lesssim1.8$ eV~cm$^{-3}$ at the solar circle \citep[e.g.][]{Boulares1990,Webber1998,Cummings2016}. By contrast, the CRdiff and CRadv models show much higher CR energy densities.

We have extensively compared and discussed the gamma-ray luminosities of our model galaxies in comparison to observations in Fig.\ \ref{fig:gamma}. Above a SFR of $\sim 0.35$ \Msun yr$^{-1}$ (comparable to that of M31) our CRdiffalfven model is in excellent agreement with the Fermi observations in the energy band $0.1-100$~GeV. By contrast, the other two models show much higher gamma-ray luminosities and over-predict the observed values. Below a SFR of $0.35$ \Msun yr$^{-1}$ there are only two data points for the SMC and LMC, which have lower gamma-ray luminosities than our models. As discussed above, the SMC and LMC are satellites of the MW which might either have enhanced SFRs in comparison to non-satellite dwarf galaxies  \citep{Harris2009} effectively shifting them to the right in Fig.\ \ref{fig:gamma} or show lower gamma-ray luminosities in comparison to the observational relation of \citet{Rojas-Bravo2016} due to an increased gamma-ray variance at low SFRs. Therefore, the paucity of gamma-ray data at low SFRs precludes strong conclusions about the correctness of models and shows the importance of obtaining better observational gamma-ray constraints at low SFRs.

We conclude that the CR energy density at the solar circle and the gamma-ray luminosities in our CRdiffalfven model reproduce the observational data well while our CRdiff and CRadv models overpredict both observational constraints.

\subsection{CR diffusion coefficient}
\label{sec:kappa}

The pressure-carrying CR distribution at GeV energies propagates via streaming with Alfv\'en waves and diffusion relative to the frame of those waves \citep{Kulsrud1969,Amato2018,Thomas2020}. If this microscopic CR transport were modelled in the form of CR diffusion, this would imply a spatially and temporarily varying CR diffusion coefficient because of the varying Alfv\'en speed and the various wave damping mechanisms that dominate in different phases of the ISM and modulate the coupling strength of CRs to the ambient plasma \citep{Jiang2018,Thomas2019}: weak wave damping implies a strong coupling and causes the CRs to stream with Alfv\'en waves whereas strong wave damping leaves less waves to scatter CRs so that CR diffusion prevails. If this complex CR transport were modelled with a constant, isotropic CR diffusion coefficient, it assumes typical values for GeV CRs of $\kappa_\rmn{iso}\sim(1-3)\times10^{28}~\rmn{cm}^2~\rmn{s}^{-1}$ as we will argue below.

The relation between $\kappa_\rmn{iso}$ and the CR diffusion coefficient along the magnetic field, $\kappa_\parallel$, depends on the exact magnetic field configuration. For a turbulent field, $\kappa_\rmn{iso} = \kappa_\parallel/3$, whereas if CR transport along the homogeneous magnetic field dominates, we have $\kappa_\rmn{iso} = \kappa_\parallel$. For CR feedback to be active, CRs move along open field lines from the disc into the halo (that are either vertically aligned through an outflow or via the Parker instability) so that $\kappa_\rmn{iso} \sim \kappa_\parallel$ in the regime of interest where galactic winds are accelerated by the CR pressure gradient. There is evidence that CRs escaping from the disc into the halo excite the streaming instability \citep{Kulsrud1969,Evoli2018}, which limits the drift speed to that of the Alfv{\'e}n frame, which is $\sim30\mbox{km s}^{-1}$ at the disc-halo interface and increases to $\sim 300\mbox{km s}^{-1}$ at the virial radius. This coincides with the diffusion velocity $\varv_\rmn{diff}\sim30\mbox{ km s}^{-1} \kappa_{28}\, L_{3\,\rmn{kpc}}$, where $\kappa_{28}=10^{28}\mbox{ cm}^2\mbox{ s}^{-1}$ is the CR diffusion coefficient and $L_{3\,\rmn{kpc}}$ is the CR gradient length, justifying our choice of the CR diffusion coefficient.

A diffusion coefficient for CRs in the Galactic disc can also be estimated from CR propagation models and observations of synchrotron radiation and/or the ratio of secondary to primary nuclei \citep{Strong1998,Ptuskin2006,Ackermann2012,Tabatabaei2013,Amato2018}. In the Galactic halo, CRs have a scale height of $\sim3$ kpc and their residency time in the thick disc is inferred to be $\tau\sim3\times10^7$~yr as obtained from measurements of the ratio of secondary-to-primary CR nuclei \citep{Lipari2014,Evoli2020}. Thus, the diffusion coefficient of GeV CRs is given by $\kappa_\rmn{iso}\sim H^2/(3\tau)\sim3\times10^{28}~\rmn{cm}^2~\rmn{s}^{-1}$. This order of magnitude estimate for $\kappa_\rmn{iso}$ is confirmed by several different studies of CR propagation, including GALPROP simulations that aim to reproduce the Fermi gamma-ray sky \citep{Porter2017,Johannesson2019}. The recently discovered hardening of the momentum power-law slope of the CR proton spectrum at low Galactocentric radii could be a signature of anisotropic diffusion in the complex Galactic magnetic field with $\kappa_\parallel=1\times10^{28}\mbox{ cm}^2\mbox{ s}^{-1}$, as suggested by DRAGON2 simulations \citep{Cerri2017,Evoli2017}. Finally, the flux of unstable secondary CR nuclei in the recent AMS-02 data, produced by spallation processes in the ISM, can be used to constrain the residence time of CR inside the Galaxy, yielding identical values for the diffusion coefficient \citep{Evoli2019,Evoli2020}.

The exact numerical value of $\kappa_\parallel$ determines the CR diffusion timescale in the disc and thus controls the amount of dynamical impact of CRs on the galaxy. A large value of $\kappa_\parallel$ leads to short diffusion timescales and a quick escape of CRs from the disc. We have tested the impact of a three times higher value of the CR diffusion coefficient ($\kappa_\parallel=3\times10^{28}~\rmn{cm}^2~\rmn{s}^{-1}$) on the results obtained here. Figure~\ref{fig:high_kappa} shows from top to bottom face-on and edge-on projections of the stellar disc's light, the gas surface density and edge-on projections of the CGM CR-to-thermal pressure ratio, $X_{\rm cr}$ for the fiducial models (with $\kappa_\parallel=1\times10^{28}~\rmn{cm}^2~\rmn{s}^{-1}$) and those with a three times higher diffusion coefficient. We confirm the expectation that a higher value of $\kappa_\parallel$ leads to a smaller dynamical effect of CRs on galaxy properties such as the reduction in stellar disc size in the CRdiffalfven model (upper panel in Fig.~\ref{fig:high_kappa}). Here the faster CR transport also causes faster Alfv\'en cooling which diminishes the dynamical impact of CRs so that the gas distribution looks similar to the noCR model (cf.\ Figs.~\ref{fig:rgb} and \ref{fig:gas}). This effect is less pronounced in the CRdiff model as here the combination of CR energy conservation (during the diffusion step) and their higher escape speed lead to a larger CR pressure-dominated halo (bottom right panels of Fig.~\ref{fig:high_kappa}). This CR pressure dominated halo in turn prevents the gas from efficiently cooling onto the disc. These simulations reinforce our main results that small variants of CR transport can have substantial impact on the resulting galaxies.

\subsection{Implications for CR transport in galaxies and the CGM}
\label{subsec:dis}

Our analyses of CR feedback have several important implications for CR transport and the excitation of CR driven instabilities. The CR pressure in the CRdiff model dominates over the thermal pressure up to large radii ($r\lesssim80$~kpc). In such a quasi-hydrostatic atmosphere the CGM necessarily attains a comparably smooth distribution. By contrast, the CR pressure distribution in the CRadv model reflects the dominating modes of transport and cooling processes. Advection of CRs with the galactic outflows along streamlines implies a highly structured CR distribution (Fig.~\ref{fig:CGMgas}). Turbulent mixing in the CGM \citep{Pakmor2019} causes a smoother CR distribution, in particular at large radii $r\gtrsim50$~kpc. 

Additionally including CR diffusion smooths the CR distribution considerably as a result of two effects: (i) CR diffusion on cosmological timescales results in a root mean square displacement of $25~\rmn{kpc}~\sqrt{\kappa_{28} t_{10~\rmn{Gyr}}}$ along the magnetic field lines and (ii) perpendicular transport is achieved through field line wandering. Assuming that the velocity differences between neighboring points follow a Gaussian distribution, we obtain explosive Richardson diffusion with a displacement $\langle x^2\rangle \propto t^3$ up to the injection scale of turbulence (and standard diffusion above this scale), which smooths the CR distribution in the CGM considerably (Fig.~\ref{fig:CGMgas}).

Most surprisingly, by additionally accounting for CR Alfv\'en wave losses, the CR pressure distribution becomes highly structured. As CRs diffuse a scale height, they loose an e-folding of their initial energy. This distance can be substantially increased if CRs are predominantly advected with the gas. As their diffusive transport reaches the effective scale height, they cool quickly and subsequent turbulent mixing is greatly suppressed. Hence they should trace out individual streamlines of the gas in their pressure as well as in the CR-to-thermal pressure ratio (see Fig.~\ref{fig:CGMgas}).

We have seen in Fig.~\ref{fig:pressure} that the CR and magnetic pressures vary by four orders of magnitude but trace each other within a factor of five out to the virial radius. This remarkable finding has severe consequences for the existence of current driven CR instabilities. The condition for exciting the hybrid, non-resonant CR instability is $\epsilon_\CR/\epsilon_B \gtrsim 2 c / \varv_\rmn{d}$ \citep{Bell2004}, where $\varv_\rmn{d}$ is the drift speed of CRs that is close to the Alfv\'en speed as explained above. Because our CR and magnetic energy densities closely trace each other, the Bell instability is not excited and hence, no additional growth of the magnetic field is expected from this plasma effect. This also implies that the adopted diffusion coefficient remains valid and is not lowered to the classical Bohm limit due to strong Bell fluctuations, which would scatter CRs off magnetic irregularities at every gyro orbit. This fast CR scattering would manifest itself as a much reduced CR diffusion coefficient by about seven orders of magnitudes, which would effectively imply a transition to the CRadv model.

\subsection{Comparison to previous work}
\subsubsection{Effects on the central galaxy}

In this study we found that CRs have little effect on global galaxy properties such as stellar mass and SFR. All our galaxies exhibit a rotationally supported gas disc dominated by the kinetic pressure (Fig.\ \ref{fig:pressure}) and a mostly thermally supported CGM in the halo ($R\gtrsim75$ kpc). The CRdiff run additionally shows a transition region at the disc-CGM interface ($20<R<75$ kpc) where CRs dominate the pressure budget, permitting a lower CGM temperature \citep[see also][figure 6]{Butsky2018}. On the other hand, in the CRadv run this transition region shows an equilibrium of CR pressure with the thermal and the kinetic pressure. This agrees qualitatively with results obtained by \citet[][figure 2]{Salem2016} and the recent findings of \citet{Hopkins2019} despite the large differences in the diffusion coefficients used (this work: $\kappa_\parallel=1\times10^{28}$ cm$^{2}$ s$^{-1}$ vs. ``best-fit'' $\kappa_\parallel=3\times10^{29}$ cm$^{2}$ s$^{-1}$ in FIRE-2). 

While qualitative agreement between our results and the ones presented by the FIRE group exist, the fundamental differences in the value of the diffusion coefficient is a result of the physical consequences of a different treatment of the ISM in the simulations. As detailed in Sect.~\ref{sec:kappa}, our choice for $\kappa_\parallel$ is justified by CR propagation studies \citep{Porter2017,Cerri2017,Evoli2017,Evoli2019,Evoli2020,Johannesson2019} and is in line with other simulation analyses of galaxies forming in a cosmological environment \citep{Salem2016}. By contrast, the favoured diffusion coefficients in the FIRE-2 simulations are a factor 10-30 larger than those studied here and inferred in the CR literature. 
In fact, we believe that the choice of such a large diffusion coefficient in the FIRE-2 simulations follows from the particular multi-phase ISM model used there and the associated difficulty to accurately model the appropriate wave damping in the different phases of the ISM, as we lay out below.

In order for the CRs to escape the dense gas without significant hadronic losses, the diffusive timescale $\tau_\rmn{diff} \approx L^2/\kappa_\parallel$ needs to be smaller than the hadronic loss time $\tau_\rmn{had} \approx 1 / (n \sigma_\rmn{pp} c)$, where $L$ is the scale height of the disc, $\sigma_\rmn{pp}\approx25$~mbarn is the total inelastic proton-proton cross section at kinetic proton energies of $1-3$~GeV \citep{Kafexhiu2014}, and $c$ is the light speed. By rearranging this inequality we derive a lower bound on the effective diffusion coefficient needed to allow the CRs to leave the galaxy, 
\begin{align}
    \label{eq:kappa}
    \kappa_\parallel>L^2n\sigma_\rmn{pp} c
\end{align}
In the AURIGA model the star forming phase is governed by an equation of state \citep{Springel2003} resulting in a relatively smooth gas distribution where stars form at gas densities above $n_{\rm th}\sim0.13$ cm$^{-3}$. This results in flat and extended gas disks in the AURIGA simulations (top panels of Fig.~\ref{fig:gas}) of scale height $\sim1.5$ kpc (see vertical gas profiles in Fig.~\ref{fig:vert_prof}) with densities in the disk mid-plane of $n\lesssim 0.4$ cm$^{-3}$ (cf. left panels in Fig.~\ref{fig:prof}). Using this density in Eqn.~\eqref{eq:kappa} shows that for the AURIGA model a value of  $\kappa_\parallel\gsim10^{28}$ cm$^{2}$ s$^{-1}$ allows CRs to escape.

On the other hand, in the FIRE-2 simulations the ISM is treated differently and allows for the formation of a multiphase ISM including a cool and dense phase with a significant amount of gas at densities $n\gsim10$ cm$^{-3}$ \citep[see Fig. 9 of][]{Hopkins2019} in their $L_\star$ galaxies. For those densities and assuming similar gas disk scale heights, Eqn.~\eqref{eq:kappa} results in $\kappa_\parallel\gsim2\times10^{29}$ cm$^{2}$ s$^{-1}$ in order to allow the CRs to escape efficiently from the ISM.
Physically, ion-neutral damping would strongly damp the self-excited Alfv\'en waves so that CRs become weakly coupled to the largely neutral gas and could escape almost ballistically at their intrinsic speed of light from these regions \citep{Wiener2017}. Once they enter the warm-hot phase of the ISM, they couple again to the gas (because of the weaker wave damping processes such as non-linear Landau damping) so that they are transported at the Alfv\'en wave speed, which corresponds to an effective parallel diffusion coefficients in the MW today of $\kappa_\parallel=(1-3)\times10^{28}$ cm$^{2}$ s$^{-1}$. A failure to model this transition results in a more diffusively transported CR gas in the warm-hot phases of the ISM and CGM \citep[see also discussion in][]{Salem2014b}. It has been suggested by \citet{Farber2018} that the decoupling of CRs at low temperatures and high densities can be approximated by artificially increasing the diffusion coefficient in low temperature gas by a factor of 30 in comparison to the general CGM gas (their equation~13). This work shows that with this simple decoupling mechanism CRs can efficiently escape the ISM and their dynamical effect is dramatically increased.

The strongest impact of CRs on the central galaxy we can find in our study is the reduction in size of the gaseous and stellar disc. Very similar results regarding the sizes of the stellar discs have been obtained by the FIRE-2 group \citep[see e.g. their Fig.\ 12 in][]{Hopkins2019} who did not investigate the underlying reasons. Here we find that the interplay of CRs and the wind feedback model on the scales of the gas disc mediates large scale effects on the accretion flow of gas. In the CR runs we observe more spherically symmetric outflows blocking the coherent gas accretion in the direction of the disc plane. This in turn modifies the angular momentum acquisition of the galaxy and manifests itself in accretion of gas from a smaller region.

Our results are in stark contrast to the results from \citet{Salem2014b} where the stellar disc grows in size when CRs are considered. The reason for this is not entirely clear but the very different feedback implementations and resolution effects might certainly play a role. These earlier results analysed simulations of worse resolution compared to the ones used here. Furthermore, the fiducial model used in that study results in a very compact stellar disc of unrealistic size. In contrast to this, the AURIGA-CR models start out from galactic discs of realistic size and mass because the wind feedback of the AURIGA noCR runs was tuned to reproduce MW-like galaxies. We discuss the uncertainties of the wind model in more detail in the next section.

\subsubsection{Effects on the circum-galactic medium} 
The most noticeable effect of CRs in the AURIGA simulations is on the structure and morphology of the CGM. Whereas the overall baryonic mass in the CGM is not drastically effected (see density profiles in Fig.\ \ref{fig:CGMprof}), the additional CR pressure affects the small scale density and temperature distribution of the CGM. In particular our CRdiff model has a smoother and cooler CGM which is maintained by the additional CR pressure which smoothes out small-scale high density clumps in the CGM and supports the gas at lower temperatures against gravity. These findings are qualitatively similar to earlier results presented in the literature \citep[e.g.][]{Salem2014b,Salem2016,Chen2016,Butsky2018}.

The findings that the CGM becomes smoother and slightly cooler when diffusing CRs are included is consistent with the results of small scale ISM simulations of the galactic disc using stratified boxes \citep{Girichidis2016,Girichidis2018,Simpson2016}. In particular, \citet{Girichidis2018} finds that the CR pressure inside the disc ($z\lesssim1$ kpc) is largely in equilibrium with the thermal pressure as is the case for all our CR runs (compare Fig.\ \ref{fig:gas}). At distances larger than that ($z\gtrsim1$ kpc), the CR pressure starts to dominate over the thermal pressure with values of $P_{\rm cr}/P_{\rm th}=10-100$ in good agreement with our results. Thus, despite the approximations of our ISM model and the comparatively lower resolution results on the scales studied here, the simulations appear to be converged.

By contrast, our CRdiffalfven model shows a warmer CGM in comparison to the model without CRs (noCR) and in strong contrast to earlier work presented in the literature \citep[e.g.][]{Salem2014b,Salem2016,Chen2016}. As explained above, the reason is the additional CR Alfv\'en wave cooling term that emulates CR energy losses as they are resonantly exciting Alfv\'en waves which scatter their pitch angles (angle between their momentum and mean magnetic field vectors). This causes them to isotropise in the Alfv\'en wave frame and to stream with the Alfv\'en velocity along the local direction of the magnetic fields \citep{Wiener2017}. Whereas this approximation is justified as long as CR streaming and diffusion fluxes match each other, this cannot be guaranteed at all times due to the dispersive mathematical nature of the diffusion operator. Clearly more work is needed to confirm this finding and to better understand the final state of the CGM in the presence of streaming CRs. On the contrary, recent results by the FIRE-2 simulations suggest that CRs are able to reduce the CGM temperature from $\gtrsim10^5$ K to $\sim10^4$ K \citep[see figure 7 in][]{Ji2019} by providing enough pressure support. Our simulations do not support such a drastic change in CGM temperature as we have shown in Fig.\ \ref{fig:phase}. In fact, a complete suppression of the hot phase as in the FIRE-2 simulations is at odds with X-ray observations of the MW hot halo \citep[$\sim10^6$ K,][]{Fang2013,Faerman2017}. 

The most likely reason for this is the implementation of feedback in FIRE-2, which is very explosive and could cause a quenching of their magnetic dynamo. This yields to saturation at a low level with a magnetic energy density that is a factor of 100 below our results. Note that our magnetic field distribution matches Faraday rotation measure data of the MW and external galaxies \citep{Pakmor2018}. The lower magnetic field strength causes the Alfv\'en speed $\bvel_{\rmn{A}}=\bB/\sqrt{4\upi\rho}$ to be ten times smaller and hence, also reduces the CR Alfv\'en wave cooling rate, $|\bvel_{\rmn{A}}\bcdot\bnabla P_\CR|$, by the same factor. Hence, the FIRE-2 runs represent an extreme version of our CRdiff model, in which the CR Alfv\'en wave cooling is nearly absent.

\subsection{Modelling uncertainties}
The results obtained in this paper are subject to a number of physical modelling uncertainties which we discuss below.

\subsubsection{The AURIGA feedback model}
In this study CRs are modelled on top of the AURIGA galaxy formation model which has been calibrated to reproduce MW-like galaxies without the inclusion of CRs. We have kept any ``free'' parameter in the sub-grid model as in the AURIGA model and added the CR physics on top of this. Therefore, the comparably small impact of CR physics on global galaxy properties such as the total stellar mass or SFR (as opposed to previous findings where CRs showed strong impacts) might be due to the already efficient feedback implementation of the AURIGA model without CRs. Here, the biggest uncertainty is the effect of the wind model coupled with the CR feedback. In the AURIGA model the details of the wind model are calibrated to reproduce observed galaxy properties without the additional effects of CRs. In this study, we add CR feedback on top of the already calibrated feedback model of AURIGA without re-tuning any parameters. Whereas this strategy allows us to cleanly single out the effects of CRs, one could imagine that the calibrated AURIGA model might already account for some of the effects CRs might have on galaxy formation. Therefore, the exact choice of parameters for the wind model in combination with the effects of CRs might change the amount of angular momentum losses as observed in our study. There might exist a different combination of wind model parameters and CR feedback model in which the CGM flow is less affected by the CRs and thus the angular momentum losses are reduced. However, the cause of the different angular momentum build-up in the three CRs variants is the modified morphology of the disc-halo interface and we expect the basic effects to be robust. Nevertheless, unless the parameters of the wind model are derived from either observations or theoretical considerations the wind model presents a considerable systematic uncertainty.

\subsubsection{The ISM model}
Our simulations adopt a pressurised ISM which even in the stellar disc is relatively smooth without high density, low temperature peaks \citep[e.g.][figures 9 and 10]{Marinacci2019}. Similarly to the AURIGA feedback model we have not re-tuned any of the parameters of the multiphase model for star formation. This is justified because all our models still reproduce the Kennicutt-Schmidt relation \citep{Kennicutt1998ARA+A} against which this model was calibrated and we have explicitly checked that each of the models reproduces the normalization and slope of the observed relation. Additionally, it is not entirely clear how CRs could be included in the Springel \& Hernquist model in the first place, so that for the purposes of this study it appears most adequate to not modify parameters of the subgrid model.

Transforming to a multi-phase ISM will effect how CRs escape dense star forming regions and thus how they impact the dynamics of the ISM. Most importantly, recent modelling of CR data suggests that CRs below 200~GeV that carry most of the CR pressure are streaming with the Alfv\'en velocity and are diffusively transported at higher energies \citep{Evoli2018}. Hence, we need to model CR streaming in the self-confinement picture where CRs resonantly excite Alfv\'en waves to accurately model their transport in galaxies and the CGM using the two-moment method \citep{Thomas2019}. Following the evolution equation of small-scale resonant Alfv\'en wave energies provides a means to self-consistently model CR diffusion in the Alfv\'en wave frame. This will enable us to simultaneously account for the weaker coupling of CRs in the cold phase ($T<10^4$~K) due to increased ion-neutral damping and the stronger dynamical coupling in the warm-hot phases due to the prevalent non-linear Landau damping \citep[e.g.][]{McKenzie1983} and turbulent damping processes \citep[e.g.][]{Farmer2004,Yan2004}.

\subsubsection{The CR transport models}
Another fundamental uncertainty is given by the details of the CR transport physics and its numerical implementation. To explore the influence of CR transport on galaxy formation we decided to adopt three different variants of CR transport and focused our analysis on the question of how each variant impacts the stellar and gaseous properties. These models result in qualitatively similar global trends, but show that structural properties differ between each of the CR variants. As expected, our CRdiffalfven model that emulates CR streaming gave the most realistic results in terms of stellar and gaseous disc properties as well as for the CGM in agreement with recent findings by \citet{Butsky2018}. In particular, the resulting gamma-ray emission (see Fig. \ref{fig:gamma}) appeared to be an important discriminant of the studied CR models.

\subsubsection{Cosmological variance of accretion histories}
In this study we have focused on analysing the effects of CR physics in cosmological simulations. Additionally to the different CR transport physics the two galaxies in these kind of simulations are further affected by the different accretion histories. For example, AuL8 undergoes a major merger at a lookback time of $\sim6$ Gyr whereas Au6 has a very quiet merger history at low redshift. Thus, there are natural differences in the evolution and properties between the two haloes complicating the separation of the effects of CRs and cosmological accretion history. On the other hand, CRs do not only affect the main galaxy but also the merging satellites and thus a complete picture of their effects can only be gained by studying a large cosmological volume, which samples the complete galaxy population.

\subsubsection{Numerical resolution study}
Convergence of galaxy properties across different levels of numerical resolution is difficult to achieve in galaxy formation simulations and poses an additional challenge in understanding the physics of galaxy formation. Ideally, the outcome of a simulation should only depend on the modelled physics and not on numerical resolution. In section 6 of \citet{Grand2017} it has been shown that our baseline model, the AURIGA model without CRs, is numerically well converged. We have run additional 8 simulations with a factor of 8 and 16 lower in mass resolution in order to test the numerical robustness of our results. While we detail the resolution dependence of our results in Appendix~\ref{sec:res}, here we summarise the main results: stellar and halo masses of the central galaxies are well converged across different resolution levels (see Table~\ref{tab:res}) and we have verified that all our results and conclusions do not depend on resolution. Especially our main findings are numerically converged: those include the more compact stellar discs (see Fig.~\ref{fig:res_test} in the Appendix) mediated by the modified accretion flow in the CR runs as well as the gas discs inflated by CR pressure and the smoother CGM in the CRadv and CRdiff models in comparison to the other two models (cf.\ Fig.~\ref{fig:gas_res} and the upper row of Fig.~\ref{fig:CGMgas}). Thus, our simulations are well suited to study the effects of CRs in cosmological simulations as the evolution of the galaxies only depends on our physical modelling and not on numerical resolution.

\section{Conclusions} \label{sec:conc}

In this work we set out to study the effects of CRs on the formation of MW-like galaxies in a cosmological context. To this extend we have performed eight magnetohydrodynamical simulations in the context of the AURIGA project \citep{Grand2017} with three different models of varying complexity for the physics of CR transport. All simulations are performed with the second-order accurate moving mesh code Arepo \citep{Springel2010,Pakmor2016c} for magneto-hydrodynamics. The galaxy formation model includes detailed models for gas cooling and heating, star formation as well as stellar and AGN feedback. Additionally, the simulations include the following CR physics: the simplest model advects CRs with the gas flow (CRadv), a more complex variant additionally follows the anisotropic diffusion of CRs parallel to the magnetic field (CRdiff) while in the most complex model CRs are further allowed to cool via the excitation of Alfv\'en waves (CRdiffalfven), attempting to emulate the transport process of self-confined CR streaming.

We have studied in detail the properties of the central galaxy and the CGM and compared model predictions from the CR runs to the fiducial AURIGA model. Bulk galaxy properties are only weakly affected by CRs, whereas the morphology and angular momentum distribution of our galaxies as well as the properties of the CGM are sensitive to the details of the CR physics implementation. Our conclusions are summarized as follows:

\begin{itemize}
\item Galaxy properties like the total stellar mass, SFR or gas mass are largely unaffected by CRs and stable across different physics variants. While previous works have found that CRs are able to suppress star formation in isolated galaxy simulations, our cosmological simulations show that the SFR is largely unaffected by CR feedback (see Fig.\ \ref{fig:sfr}). Note that this could be partially due to the already efficient feedback in AURIGA, which can in principle mask some of the feedback effects CRs would otherwise have.

\item Comparing structural parameters of the galaxies such as disc sizes, disc-to-total stellar mass ratios or gas disc morphology we find strong differences between the simulations that include CRs and the fiducial AURIGA model. The CRadv and CRdiff models result in more compact, bulge dominated discs which show thicker and smoother gas discs, in which the vertical force balance is dominated by the CR pressure (see e.g., Figs.\ \ref{fig:rgb} and \ref{fig:gas}). A similar reduction of stellar disc size is also found by the FIRE-2 group in their simulations including CR feedback. In contrast, the stellar and gaseous discs in the CRdiffalfven model have disc sizes which lie between the fiducial results and the more extreme CR models (e.g., the left panel of Fig.\ \ref{fig:prof}). We find that the magnetic field strength and morphology is similar in all our runs with a value of the order of $\sim10\, \mu$G \citep[that is consistent with MW observations, see][]{Pakmor2018} so that our magnetic energy density is roughly 100 times larger than those obtained with the FIRE-2 simulations \citep[e.g. Figs. 3 and 19 of][]{Hopkins2019}.

\item The interplay of CRs and the wind feedback model strongly affects the gas flow patterns in the CGM (Figs.\ \ref{fig:flow} and \ref{fig:flow2}). The more compact, bulge dominated discs in the CR simulations cause the outflows to become more spherically symmetric in comparison to the fiducial AURIGA run (Fig. \ref{fig:flow}) and thus alter the angular momentum acquisition in the cosmological runs. In this way the action of CR feedback in the star forming disc changes the outflow geometry and suppresses the baryonic accretion of high angular momentum gas, especially at late cosmic times (Fig.\ \ref{fig:ang_mom}). As a consequence, the gas discs in the CR runs are smaller in size as is highlighted in the left panels of Fig.\ \ref{fig:prof}.

\item On larger scales, CRs strongly affect the properties of the CGM. The advection and diffusion models exhibit a smoother and partly cooler CGM (Figs.\ \ref{fig:CGMgas} and \ref{fig:CGMprof}) where the additional CR pressure is able to stabilise the CGM against gravitational collapse compensating for the missing thermal pressure support at lower CGM temperatures. These runs therefore show large ($R\sim50$ kpc) CR pressure contributions in the haloes. In contrast, the Alfv\'en wave model is only CR pressure dominated at the disc-halo interface and the CRs come into equilibrium with the thermal pressure as they are advected into the halo along stream lines of the galactic winds (see also Fig.\ \ref{fig:pressure}). As CRs are actively transported across an effective scale height, they quickly cool, which greatly suppresses further turbulent mixing and causes a highly structured CR pressure distribution in the CGM (Fig. \ref{fig:CGMgas}). This in turn causes a structured density and temperature distribution in the CGM, which maintains large volumes at thermally unstable temperatures of $10^5$~K (Fig. \ref{fig:phase}) which is warmer than the cool ($\sim10^4-10^5$ K) CGM gas found in the CR FIRE-2 simulations \citep[see Fig. 7 in][]{Ji2019}.

\item The magnetic and CR pressures trace each other within a factor of five out to the virial radius (Fig.~\ref{fig:pressure}). This implies that there is not enough free energy available to drive the hybrid, non-resonant CR instability \citep{Bell2004}, which would require the CR-to-magnetic energy density ratio to be larger than $2c/\varv_\rmn{d}\sim10^3-10^4$ where $\varv_d$ is the drift speed of CRs that is close to the Alfv\'en speed. Excitation of the Bell instability would imply fast CR scattering, a much reduced CR diffusion coefficient by about seven orders of magnitudes and effectively transition to the CRadv model.

\item There are active ongoing efforts in developing efficient and accurate CR magneto-hydrodynamical schemes \citep{Jiang2018,Thomas2019} to compute the CR feedback effects in cosmological simulations. To this end, direct observables are invaluable in constraining effective CR transport models, provided the approximations used for CR transport and the ISM are commensurate and not inconsistent. In Fig.\ \ref{fig:gamma} we compare the gamma-ray luminosity from hadronic CR interactions with the ISM of our models to observations of local galaxies. We find that the CRdiffalfven run agrees well with observed relations whereas the CRdiff and CRadv produce higher gamma-ray luminosities at the MW mass scale compared to observations. Our comparison here presents a first step towards understanding the effects of CRs on cosmological galaxy formation, but further work in this direction is needed to constrain valid CR transport coefficient and prevailing transport processes.
\end{itemize}

\section*{Acknowledgments}
We thank the anonymous referee for a careful reading of the manuscript which helped to improve the quality of the paper. TB and CP acknowledge support by the European Research Council under ERC-CoG grant CRAGSMAN-646955. This research was supported in part by the National Science Foundation under Grant No. NSF PHY-1748958. This research made use of the following {\sc{python}} packages: {\sc{matplotlib}} \citep{matplotlib}, {\sc{SciPy}} \citep{scipy}, {\sc{NumPy}} \citep{numpy},  {\sc{IPython and Jupyter}} \citep{ipython,jupyter}.

\section*{Data availability}
The data underlying this article will be shared on reasonable request to the corresponding author.



\bibliography{astro-ph.bib}

\appendix

\section{Surface density fits}

\begin{figure*}
\begin{center}
\vspace{-.35cm}
\includegraphics[width=\textwidth]{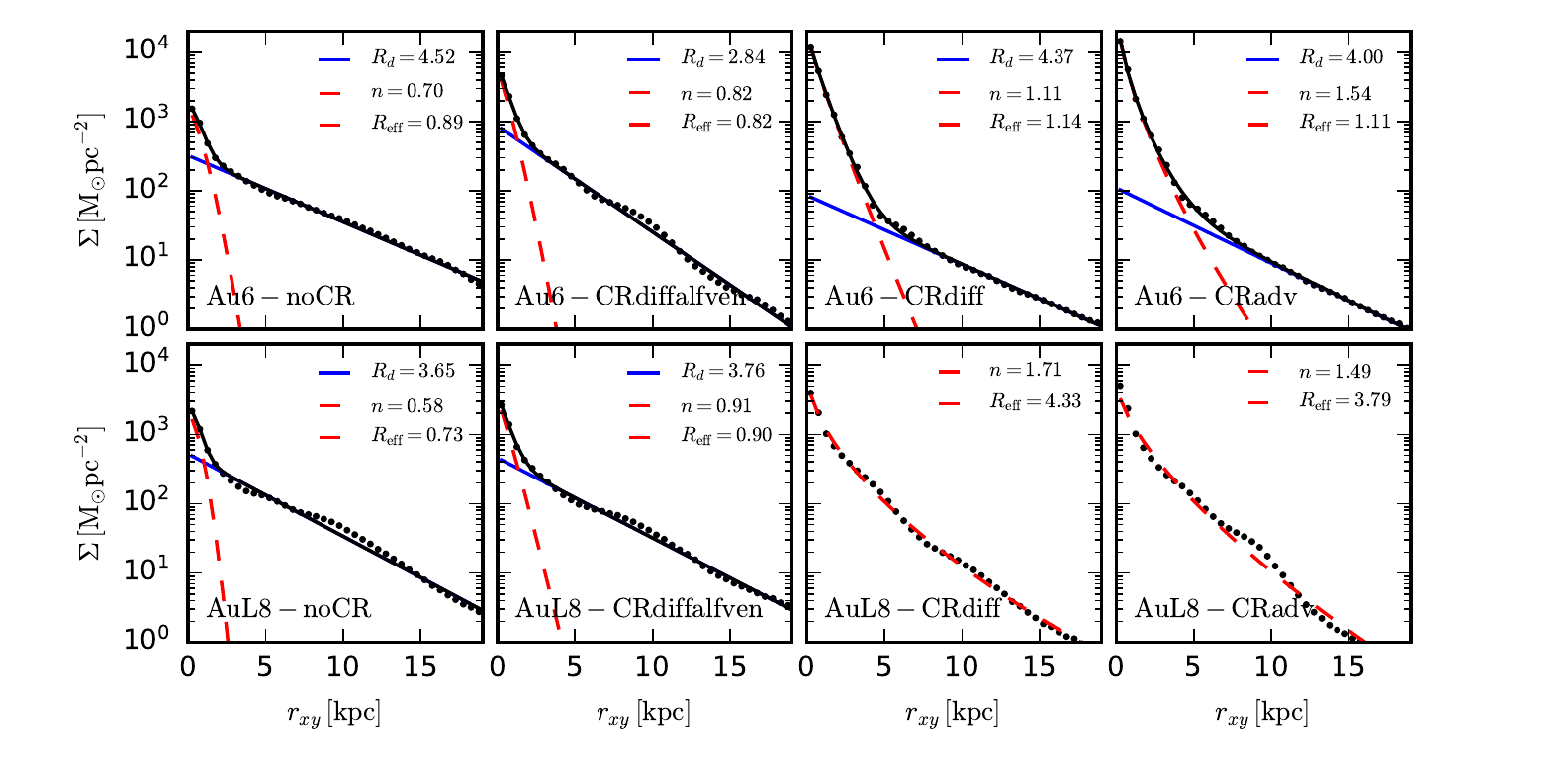}
\end{center}
\vspace{-.75cm}
\caption{Face-on stellar surface density profiles for all simulations at $z = 0$ (black dots). The four models of Au6 are shown in the upper row, AuL8 in the bottom row. The profiles are simultaneously fit with a \citet{Sersic1963} (red dashed curve) and exponential (blue curve) profile. The total fitted profile is indicated by the black curve. Resulting fit values for the disc scale length, $R_{\rm d}$, the bulge effective radius, $R_{\rm eff}$ and the bulge S\'ersic index, $n$, are given in each panel. The CRdiff and CRadv models of AuL8 are well fitted by a pure S\'ersic profile.}
\label{fig:surf_den_fit}
\end{figure*}

In Fig. \ref{fig:surf_den_fit} we show azimuthally averaged surface density fits to the stellar disc of the eight simulations at redshift $z=0$. Surface density profiles are created for all the stellar mass within $\pm5$ kpc of the mid plane in the vertical direction. The profiles are simultaneously fit with a \citet{Sersic1963} (red dashed curve) and exponential (blue curve) profile using a non-linear least squares method. Resulting fit values for the disc scale length, $R_{\rm d}$, the bulge effective radius, $R_{\rm eff}$ and the bulge S\'ersic index, $n$, are given in each panel. This figure shows that the CRadv and CRdiff runs result in more compact bulge dominated galaxies whereas the CRdiffalfven run results in a disc dominated galaxy more similar to the fiducial AURIGA run. From the fits we derive disc-to-total mass ratios (D/T) which are given in Table \ref{tab:props} in the main text.

\section{Vertical profiles}

\begin{figure*}
\begin{center}
\vspace{-.25cm}
\includegraphics[width=1.1\textwidth]{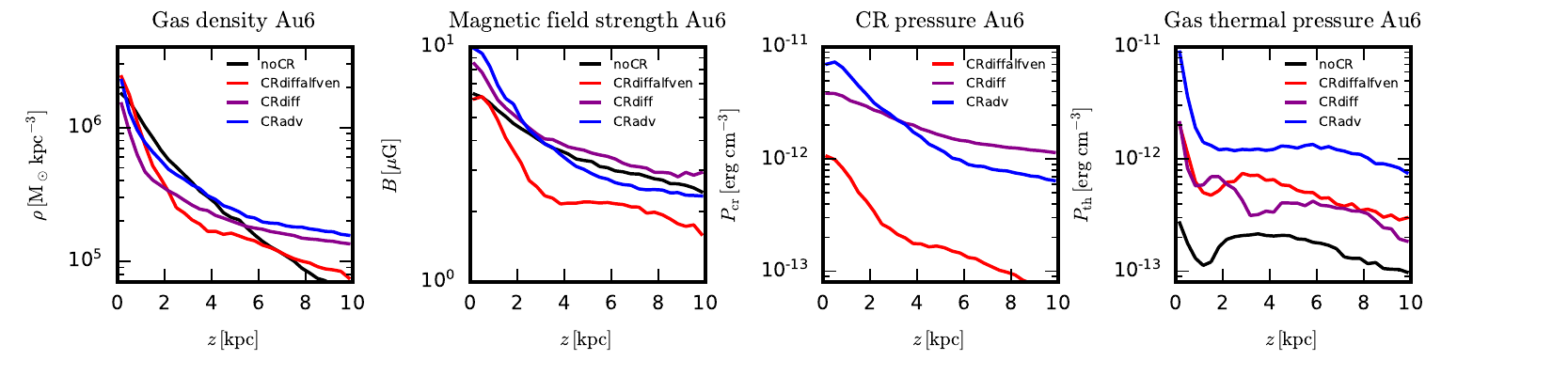}\vspace{-.25cm}
\includegraphics[width=1.1\textwidth]{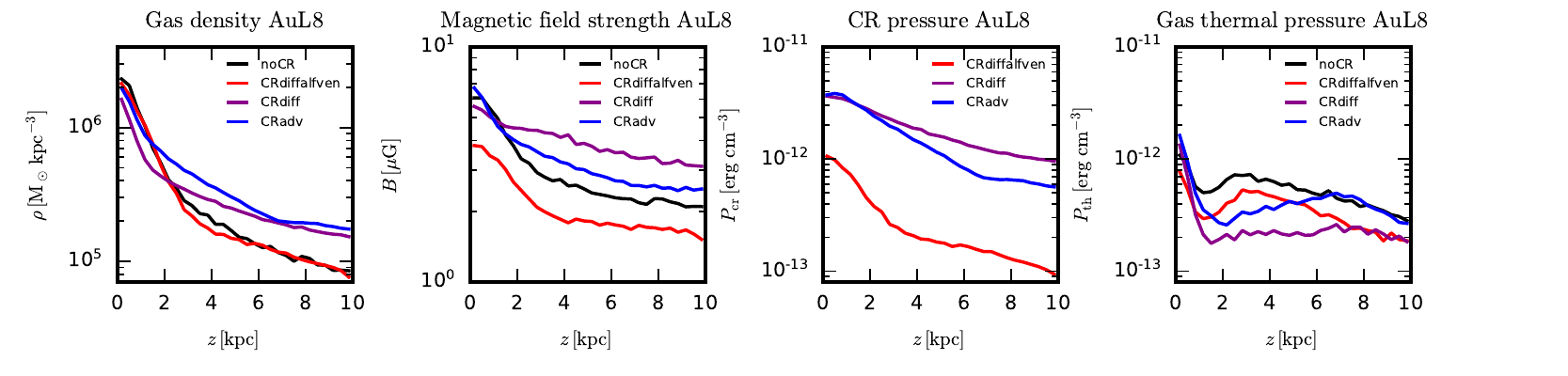}
\end{center}
\vspace{-.45cm}
\caption{Vertical profiles of the gas density (left panel), magnetic field strength (second panel), CR pressure (third panel), and gas thermal pressure
    (forth panel) for the four models of Au6 in the upper row and AuL8 in the lower row.}
\label{fig:vert_prof}
\end{figure*}

We have created vertical profiles similar to the radial profiles shown in Fig. \ref{fig:prof} for the gas density, the magnetic field strength, the CR pressure and the gas thermal pressure in the central galaxy. We select all Voronoi cells in a cylinder of radius $r=30$ kpc and height $z=\pm10$ kpc and show the data in 30 bins linearly spaced in $z$ in Fig. \ref{fig:vert_prof}.

\section{Angular momentum distribution}

\begin{figure*}
\vspace{-.35cm}
\includegraphics[width=.9\textwidth]{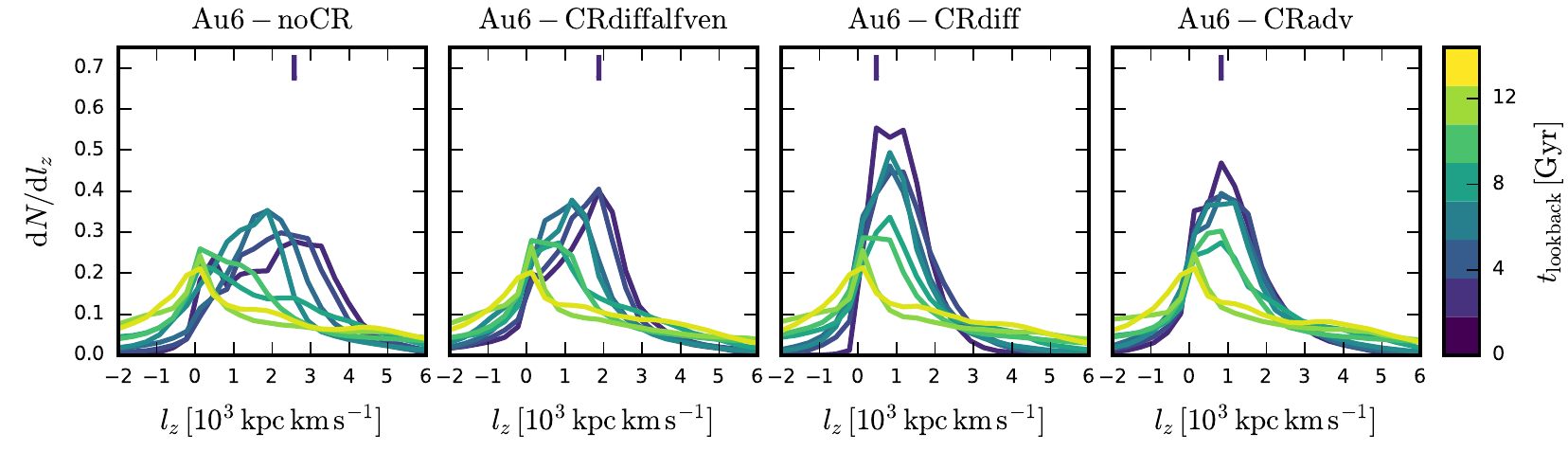}
\includegraphics[width=.9\textwidth]{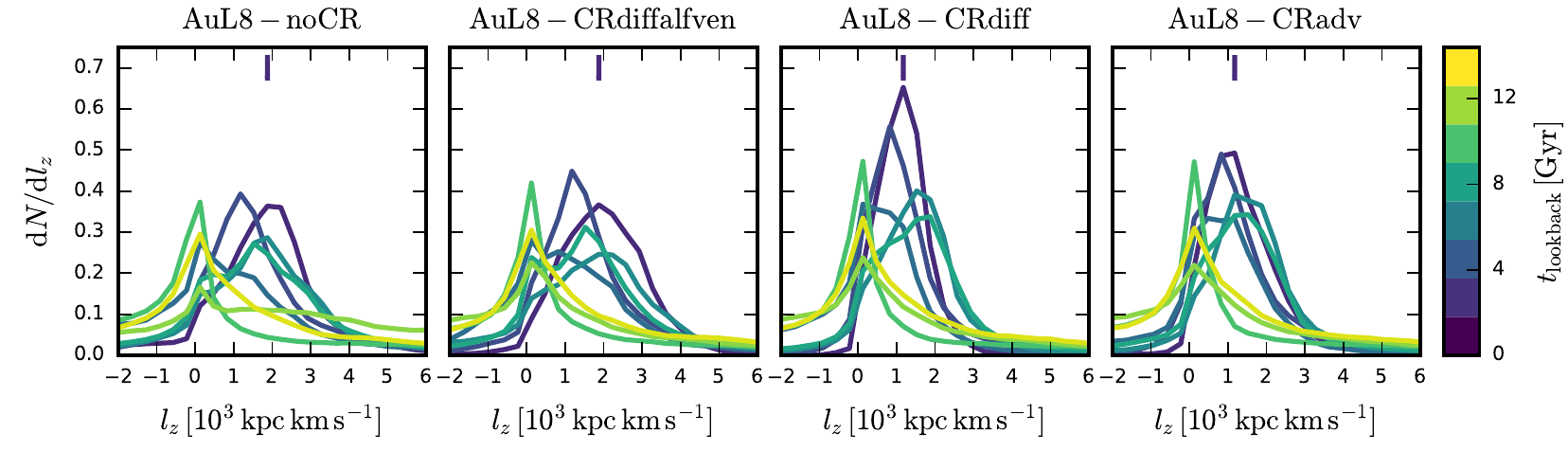}
\vspace{-.25cm}
\caption{Normalized distribution functions of the gas' specific angular momentum for different lookback times as indicated with the colorbar on the right. Upper panels show results for the four different physics variants of Au6 and lower panels for AuL8, respectively. The vertical blue line indicates the maximum of the redshift zero distribution.}
\label{fig:ang_mom2}
\end{figure*}

Figure \ref{fig:ang_mom2} shows the distribution of gas angular momentum at 8 different points in time for the tracer particles ending up in stars at present-day. This highlights how the angular momentum of the accreted gas changes over time for the different CR runs compared to the noCR run. Upper panels show galaxy Au6 and lower panels AuL8, respectively. At early cosmic times ($t_{\rm lookback}\gtrsim12$ Gyr, yellow colors) all simulations show a symmetric distribution of specific angular momenta around $l_{z}=0$ kpc km s$^{-1}$. Then, at lookback times of about $8$ Gyr (greenish colours) all runs have accreted gas with higher angular momentum of values $l_{z}\sim1.5\times10^3$ kpc km s$^{-1}$. The noCR and the CRdiffalfven runs keep acquiring high angular momentum gas also at low redshift (smaller lookback times, blue colours) whereas the angular momentum gain in the CRdiff and CRadv runs  is suppressed. Thus, at redshift zero, the angular momentum distribution in the latter cases peaks around $l_{z}\sim1\times10^3$ kpc km s$^{-1}$ and at $l_{z}\gtrsim2\times10^3$ kpc km s$^{-1}$ in the former cases (see vertical thin lines).

\section{Resolution Study}
\label{sec:res}

\begin{table}
\begin{center}
\caption{Virial mass, $M_{200}$ and stellar mass, $M_{\rm star}$ for the main galaxy across three different resolution levels for all four models.}
\label{tab:res}
\begin{tabular}{l c c c c c}
		\hline\hline
		 & & noCR & CRdiffalfven & CRdiff & CRadv \\
		\hline
		 \multicolumn{6}{c}{level 4} \\
		 \hline
		 $M_{200}$ & [$10^{12}\Msun$] & 1.02 & 1.06 & 1.07 & 1.09 \\
		$M_{\rm star}$ & [$10^{10}\Msun$] & 4.36 & 5.54 & 5.81 & 6.19 \\
		\hline
		 \multicolumn{6}{c}{level 5} \\
		 \hline
		 $M_{200}$ & [$10^{12}\Msun$] & 1.03 & 1.02 & 1.07 & 1.06 \\
		$M_{\rm star}$ & [$10^{10}\Msun$] & 4.24 & 4.19 & 5.87 & 5.80 \\
		 \hline
		 \multicolumn{6}{c}{level 6} \\
		 \hline
		 $M_{200}$ & [$10^{12}\Msun$] & 0.94 & 0.99 & 1.00 & 1.03 \\
		$M_{\rm star}$ & [$10^{10}\Msun$] & 3.78 & 2.84 & 3.44 & 3.71 \\
        \hline
\end{tabular}
\end{center}
\end{table}

\begin{figure*}
\vspace{-.45cm}
\hspace*{-.35 cm}
\includegraphics[width=.95\textwidth]{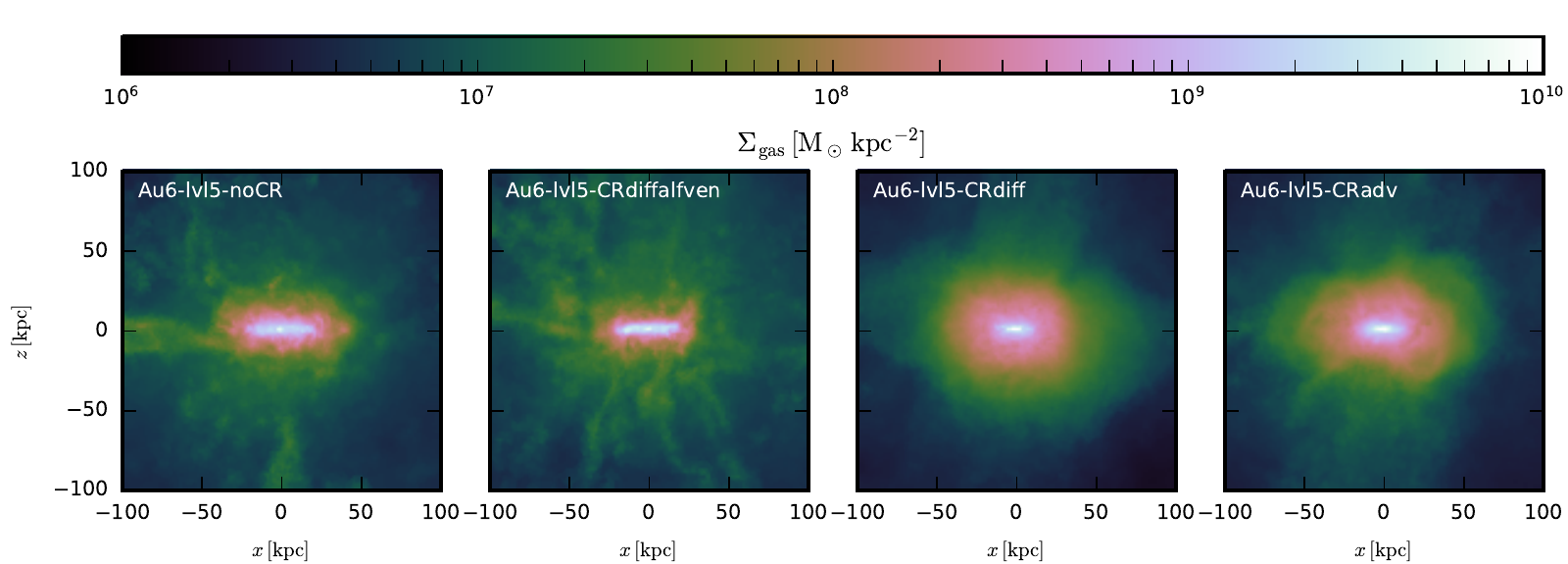}
\caption{Gas surface density maps of Au6 level 5 for all four models as indicated in the panels. Orientation and projection depth are as in Fig. \ref{fig:CGMgas}.}
\label{fig:gas_res}
\end{figure*}

\begin{figure*}
\hspace*{-1.5 cm}
\includegraphics[width=1.15\textwidth]{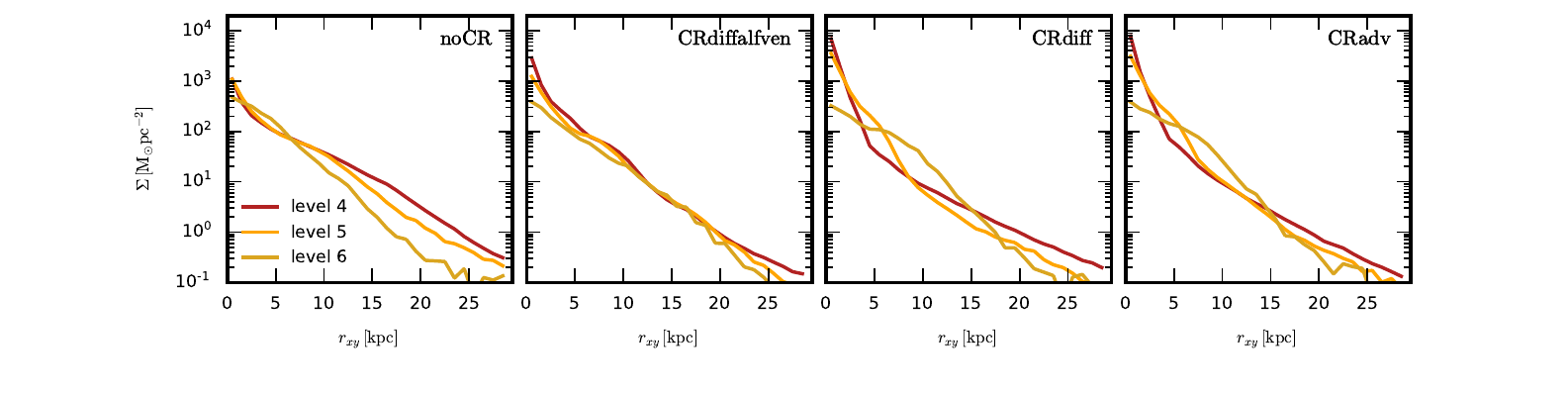}\vspace{-.5 cm}
\includegraphics[width=1.05\textwidth]{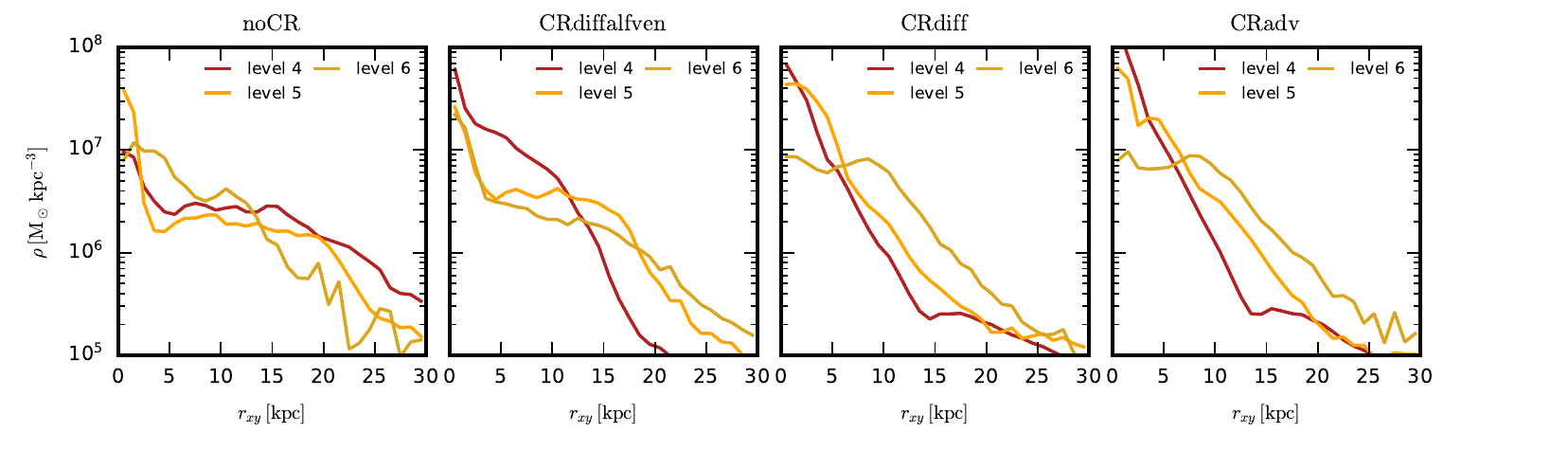}\vspace{-.15cm}
\hspace*{3.9 cm}
\includegraphics[width=.8\textwidth]{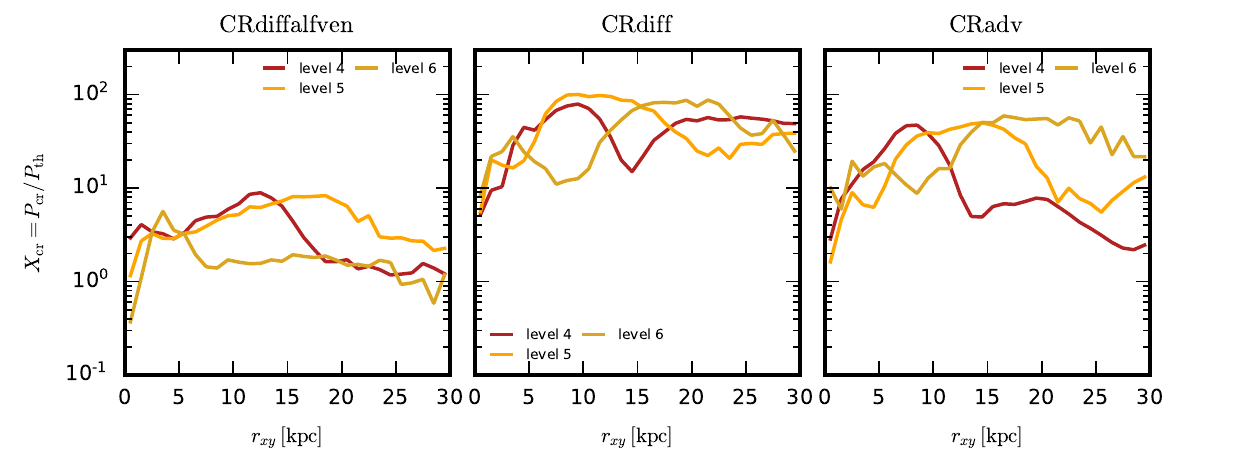}
\vspace{-.75cm}
\caption{Comparison of the profiles of stellar surface density (upper panels), gas density (middle panels) and the CR-to-thermal pressure ratio (bottom panels) of the galaxy Au6 at three different resolution levels (as indicated in the figure legends).}
\label{fig:res_test}
\end{figure*}

Our study shows that CRs strongly affect the CGM and the gaseous and stellar properties of the galactic discs. In combination with the model for the wind feedback, this results in a more hydrostatic gas halo and a modification of the gas accretion onto the central galaxy. This effect is already present at resolution levels 5 and 6 (at a factor of 8 and 16 lower in mass resolution, corresponding to $m_{\rm dm}=2\times10^6\Msun$, $m_{\rm b}=4\times10^5\Msun$, $\epsilon=738$ pc and $m_{\rm dm}=2\times10^7\Msun$, $m_{\rm b}=3\times10^6\Msun$, $\epsilon=1476$ pc) and does not change at our fiducial resolution at level 4 ($m_{\rm dm}=3\times10^5\Msun$, $m_{\rm b}=5\times10^4\Msun$, $\epsilon=369$ pc).

We would like to emphasize that we do not change any subgrid parameters for the ISM, wind feedback and CR physics when we change the numerical resolution. Hence, we do not expect to resolve new physics with increasing resolution but we aim at better resolving the poorly resolved regions at the disc-halo interface and the gas accretion and flow pattern in the CGM (i.e., we study convergence of our numerical model). For example, Fig.\ \ref{fig:gas_res} shows the gas surface density maps of the four models at resolution level 5. These are the same panels as in the upper row of Fig.\ \ref{fig:CGMgas} in the main text. The more compact and inflated gas discs in the vertical direction as well as the smoother CGM in the CRadv and CRdiff simulations are clearly visible. We find that at lower resolution (i.e., at resolution levels 5 and 6) this leads to the same hydrostatic CGM properties we have found for our fiducial resolution (level 4) in Sections \ref{sec:props} and \ref{sec:CGMprops}.

We further quantify the effects of resolution on the properties of the central galaxy such as the size and morphology of the stellar and gaseous disc. To this extent we compare in Fig.\ \ref{fig:res_test} radial profiles of the stellar surface density (upper panels), the gas mass density (middle panels) and the ratio of CR-to-thermal pressure (bottom panels) for all three resolution levels. Stellar and gaseous disc properties are in general well converged across all resolution levels (see also Table \ref{tab:res}). However, we note some differences  of the central stellar and gas density in the lowest resolution simulations (level 6) for the CRadv and CRdiff models. For the fiducial AURIGA model, on the other hand, we see that the radial density profiles of the lowest resolution simulations results are slightly steeper. In the CRdiffalfven model the stellar surface density profile is remarkably similar across all resolution levels while the gas density profiles of level 5 and 6 slightly differ from the highest resolution level. However, we note that these differences at various resolution levels are smaller than the differences found between the two haloes studied in the main text. This argues that cosmic variance causes larger differences and that our models are sufficiently numerically converged, not only for global quantities but also for all radial profiles of interest.

Most importantly for our study is that the implementation of CR physics is converged across different resolution levels. In the bottom panel of Fig.\ \ref{fig:res_test} we compare the ratio of CR-to-thermal pressure across the three resolution levels and find overall good agreement between the results. The biggest differences appear for the CRadv run where the $X_{\rm cr}$ values at large radii are higher for level 5 and level 6 in comparison to the fiducial level 4 run. 

To conclude, we find that the simulations presented here show good numerical convergence of stellar, gaseous and CR properties across three levels of resolution. This suggests that our models are well posed to study the effects of CRs on the evolution of MW-like galaxies because the simulation properties solely depend on physical parameters and not on numerical resolution.

\label{lastpage}

\end{document}